\documentclass[
10pt,
showpacs,preprintnumbers,nofootinbib,
twocolumn,
 amsmath,amssymb,
prc,
groupedaddress,superscriptaddress,
floatfix,
]
{revtex4-1}
\usepackage{graphicx}
\usepackage{bm}
\usepackage[colorlinks=true,urlcolor=blue,citecolor=blue,breaklinks=true]{hyperref}
\usepackage{epsfig}

\newcommand{\cut}[1]{}

\newcommand{\nalpha}{$n$-$\alpha$ }
\newcommand{\ntlos}{N$^3$LO$^*$ }

\begin{document}


\title{Quantifying uncertainties in neutron-alpha scattering with chiral nucleon-nucleon and three-nucleon forces}
\author{Konstantinos Kravvaris}
\email{kravvaris1@llnl.gov}
 \affiliation{Lawrence Livermore National Laboratory, P.O. Box 808, L-414, Livermore, California 94551, USA}
\author{Kevin R.\ Quinlan}
\email{quinlan5@llnl.gov}
 \affiliation{Lawrence Livermore National Laboratory, P.O. Box 808, L-414, Livermore, California 94551, USA}
\author{Sofia Quaglioni}
 \affiliation{Lawrence Livermore National Laboratory, P.O. Box 808, L-414, Livermore, California 94551, USA}
\author{Kyle A. Wendt}
 \affiliation{Lawrence Livermore National Laboratory, P.O. Box 808, L-414, Livermore, California 94551, USA}
 \author{Petr Navr\'atil}
 \affiliation{TRIUMF, 4004 Wesbrook Mall, Vancouver, British Columbia, V6T 2A3, Canada}

\date{\today}

\begin{abstract}
{
\noindent {\bf Background:} Modern \textit{ab initio} theory combined with high-quality nucleon-nucleon (NN) and three-nucleon (3N) interactions from chiral effective field theory (EFT) can provide a predictive description of low-energy light-nuclei reactions relevant for astrophysics and fusion-energy applications. However, the high cost of computations has so far impeded a complete analysis of the uncertainty budget of such calculations.
\\
{\bf Purpose:} 
Starting from NN potentials up to fifth order (N$^4$LO) combined with leading-order 3N forces, we study how the order-by-order convergence of the chiral expansion and confidence intervals for the 3N contact and contact-plus-one-pion-exchange low-energy constants 
($c_E$ and $c_D$) contribute to the overall uncertainty budget of many-body calculations of neutron-$^4$He ($n$-$\alpha$) elastic scattering.\\
{\bf Methods:} 
We compute structure and reaction observables for three-, four- and five-nucleon systems
within the \textit{ab initio} frameworks of the no-core shell model an no-core shell model with continuum.
Using a small set of design runs, we construct a Gaussian process model (GPM) that acts as a statistical emulator for the theory. 
With this, we gain insight into how uncertainties in the 3N low-energy constants propagate throughout the calculation and determine the Bayesian
posterior distribution of these parameters with Markov-Chain Monte-Carlo.\\
{\bf Results:} 
We find rapidly converging $n$-$\alpha$ phase shifts with respect to the chiral order. 
With the adopted leading-order 3N force, calculations based on the NN interaction at N$^4$LO of Entem, Machleidt and Nosyk [Phys. Rev. C {\bf 96}, 024004 (2017)] 
are unable to reproduce the experimental phase shifts in the $3/2^-$ channel 
within the estimated chiral truncation errors. Closer agreement with empirical data is found when using an older parameterization of the NN interaction at order N$^3$LO [Entem and Machledit, Phys. Rev. C {\bf 68}, 041001(R) (2017)], and the position and width of the $P$-wave resonances can be used to reduce the uncertainty of the 3N low-energy constants. \\ 
{\bf Conclusions}
The present results point to a lack of spin-orbit strength when the newer parameterization of the chiral NN force up to fifth order is combined with the leading order 3N force. The inclusion of higher-order 3N-force terms 
may be required to recover the missing strength.
GPMs can act as fast and accurate emulators of \textit{ab initio} many-body calculations of low-energy scattering and reactions of light nuclei, opening the way to a robust quantification of theoretical uncertainties grounded in the description of the underlying chiral Hamiltonian. 

}
\end{abstract}

\maketitle

\section{Introduction}
\begin{table*}
\begin{center}
\begin{ruledtabular}
\caption{\label{tab:LECsTable} Optimal values for the $c_D$ and $c_E$ LECs of the `local/non-local' 3N force (3N$_{\rm lnl}$) constrained in this work at each given order of the chiral NN interaction of Ref.~\cite{Entem2017}, as well as for the local 3N force (3N$_{\rm loc}$) of Ref.~\cite{Gazit2019} combined with the N$^3$LO interaction of Ref.~\cite{Entem2003} denoted as N$^3$LO$^{*}$. For the N$^4$LO+3N$_{\rm lnl}$ and N$^3$LO$^{*}$+3N$_{\rm loc}$ chiral interaction models  we also list the confidence intervals obtained from including the estimated chiral truncation error at that given order (see text for details). }
\begin{tabular}{ l c ccc c c c}
Chiral Models & Refs. & $c_D^\mathrm{opt}$ & $c_E^\mathrm{opt}$ & $c_D^{\mathrm{min}}$ & $c_D^{\mathrm{max}}$ & $c_E^\mathrm{min}$& $c_E^\mathrm{max}$ \\[0.2mm]
\hline
N$^2$LO+3N$_{\rm lnl}$ & &$-0.37$ & $-0.189$ & - & - & - & -\\[0.2mm]
N$^3$LO+3N$_{\rm lnl}$ & \cite{Entem2017}& $-1.33$ &  $-0.413$ & - & - & - & -\\[0.2mm] 
N$^4$LO+3N$_{\rm lnl}$ & & $-1.32$ & $-0.248$ & $-2.40$ & $-0.23$ & $-0.395$ & $-0.110$\\[0.2mm]
\hline
N$^3$LO$^*$+3N$_{\rm loc}$ & \cite{Entem2003,Gazit2019}& ~~$0.83$ & $-0.052$ & $-1.46$ & ~~$3.23$ & $-0.406$ & ~~$0.248$\\[0.2mm]
\end{tabular}
\end{ruledtabular}
\end{center}\end{table*}
In light nuclei, the most accurate predictions of nuclear properties are obtained from the direct solution of the quantum mechanical problem by combining advanced many-body methods, validated models of nuclear forces, and cutting-edge computational simulations. 
The realization of such an \textit{ab initio} description for low-energy reactions of light nuclei has been in large part enabled by the ab initio no-core shell model with continuum (NCSMC) approach~\cite{Baroni2013,Baroni2013a,Quaglioni2018} coupled with the use of nucleon-nucleon (NN) plus three-nucleon (3N) forces derived within the framework of chiral effective field theory (EFT)~\cite{Weinberg1990,Entem2003,Epelbaum2006,Epelbaum2009,Entem2017}. The NCSMC is a 
many-body approach which combines static wave functions of the many-body system and continuous \textit{microscopic-cluster} states (made of pairs or triplets of nuclei in relative motion with respect to each other) to arrive at a fully integrated description of the structure and interaction dynamics of the reacting nuclei and outgoing products. 
Chiral EFT 
links nuclear forces to the fundamental theory of quantum chromodynamics by organizing them in a systematically improvable expansion, with empirically constrained parameters that capture the physics from excluded degrees of freedom.  

While in past work we demonstrated that the NCSMC combined with chiral NN+3N forces can provide a high-fidelity description of complex nuclear reactions, such as e.g.\ the deuterium-tritium fusion~\cite{Hupin2019} or $d$-$\alpha$ scattering~\cite{Hupin2015}, a complete analysis of the uncertainty budget of such calculations had not been possible. In particular, a study of the order-by-order convergence of the chiral expansion for low-energy reactions of light nuclei and an analysis of how the uncertainty in constraining the low-energy constants (LECs) of chiral interaction models reflects on the computed scattering observables are still missing. 

Performing \textit{ab initio} many-body calculations for light-nuclei reactions in $A > 4$ systems presents a formidable computational challenge; while some calculations are feasible with modern supercomputers, the large number of samples (typically thousands) one needs in order to reach desired levels of statistical uncertainty is usually either impractical or even impossible. 
To alleviate these issues one can opt to construct models that rely on a much smaller set of design runs, while delivering a fast and accurate emulation of the response of the full 
calculation. Recently, eigenvector continuation was applied to perform a global sensitivity analysis of the binding energy and the charge radius of the $^{16}$O nucleus based on a next-to-next-to-leading-order (N$^2$LO) chiral Hamiltonian with NN+3N forces~\cite{Ekstrom2019}.
Another candidate for obtaining fast and accurate emulators of computationally expensive \textit{ab initio} calculations are Gaussian process models (GPMs)~\cite{McDonnell2015}.
In  GPMs, the error in the output prediction is modeled as a multivariate Gaussian random variable with a mean of zero and a covariance function modeled in terms of the Euclidean distance between a pair of input value vectors. Therefore, the conditional distribution of the new prediction given all the output observed from the physics model is also Gaussian with a well specified mean and covariance. This makes modeling the uncertainty associated with the predictions straightforward.

In this manuscript, we explore the use of GPMs for carrying  out a rigorous uncertainty quantification and sensitivity studies of NCSMC calculations of low-energy scattering and reactions of light nuclei based on chiral NN+3N Hamiltonians.  As a first application, we focus on the uncertainty in the determination of the contact and contact-plus-one-pion-exchange LECs ($c_E$ and $c_D$) of the 3N force and consider the elastic scattering of neutrons ($n$) on $^4$He ($\alpha$), for which the convergence of the NCSMC is well understood and under control, and which has been shown to be a magnifying glass for 3N-force effects~\cite{Nollett2007,Hupin2013,Lynn2016}.  

The paper is organized as follows. In Section~\ref{sec:approach}, we present the adopted chiral interaction models and the method used to determine the central values and confidence intervals for the 3N-force LECs $c_D$ and $c_E$,  briefly review the NCSMC approach, and introduce in detail the GPM emulator.   In Sec.~\ref{sec:results}, we present the results obtained for neutron elastic scattering on $^4$He. After discussing the convergence of the chiral expansion,  we summarize the conditions of the calculations used to construct the GPM emulator, study the sensitivity of the phase shifts to the statistical uncertainty of the 3N-force LECs $c_D$ and $c_E$, and then use the empirical values of the $^3$H ground state (g.s.) energy, $^4$He g.s.\ energy and charge radius, and the position and widths of the $n$-$\alpha$ low-lying $P$-wave resonances to extract the posterior distributions of these parameters. Conclusions and outlook are given in Sec.~\ref{sec:conclusions}. Additional material not suitable for the main text is presented in Appendix.
      
\section{Approach} 
\label{sec:approach}
\subsection{Chiral interaction models}
\label{subsec:interactions}
To study the converge of the chiral expansion, we adopt the NN potentials ranging from leading order (LO)
to fifth order of chiral expansion (N$^4$LO) of Ref.~\cite{Entem2017}, for which the same power counting scheme as well as the same cutoff procedures are applied in all orders. 
These NN potentials are combined with a 3N force at next-to-next-to-leading order (N$^2$LO)~\cite{VanKolck94} using a mixture of local~\cite{Navratil2007} and non-local regulators~\cite{EkJa15}. The local momentum cutoff is taken at $650$ MeV, while for the non-local cutoff we adopt the same value ($500$ MeV) as in the NN interaction. 
For the two-pion-exchange component of the 3N force, we adopt the effective pion-nucleon ($\pi N$) LECs recommended (at each given order of the NN potential) in Table \rm{IX} of Ref.~\cite{Entem2017}. 
The remaining LECs for the contact plus one-pion exchange ($c_D$) and contact ($c_E$) terms of the 3N force are then constrained to the binding energy and the $\beta$-decay half life of tritium following the procedure of Refs.~\cite{Gazit2009,Gazit2019}, where the weak axial-vector current is given by the
Noether current built from the axial symmetry of the chiral Lagrangian up to order N$^3$LO. 
In addition to this set of chiral forces, we also consider the NN interaction up to fourth order of Ref.~\cite{Entem2003}, again accompanied by the 3N interaction of Ref.~\cite{VanKolck94} but with a purely local regulator with $500$ MeV cutoff and the LECs as in Ref.~\cite{Gazit2019}. 
This is an interaction that we have used extensively in the past for various low-energy reaction calculations~\cite{Hupin2013,Hupin2014,Hupin2015,Calci2016,Hupin2019}, and thus we are interested to see the corresponding variance of the phase shifts with respect to the 3N force LECs. A summary of the adopted chiral interaction models and associated notation is presented in Table~\ref{tab:LECsTable}. 

To constrain the 3N-force LECs for the family of chiral NN interactions of Ref.~\cite{Entem2017}, we carry out calculations for properties of $A=3$ nuclei starting from the bare interaction. We adopt the same, large Jacobi-coordinate harmonic oscillator (HO) basis space (including up to $N_{\rm max}=40$ excitations above the ground-state configuration) as in the work of Refs.~\cite{Gazit2009,Gazit2019}.

\begin{figure}[t]
\centering
\includegraphics[width=0.475\textwidth]{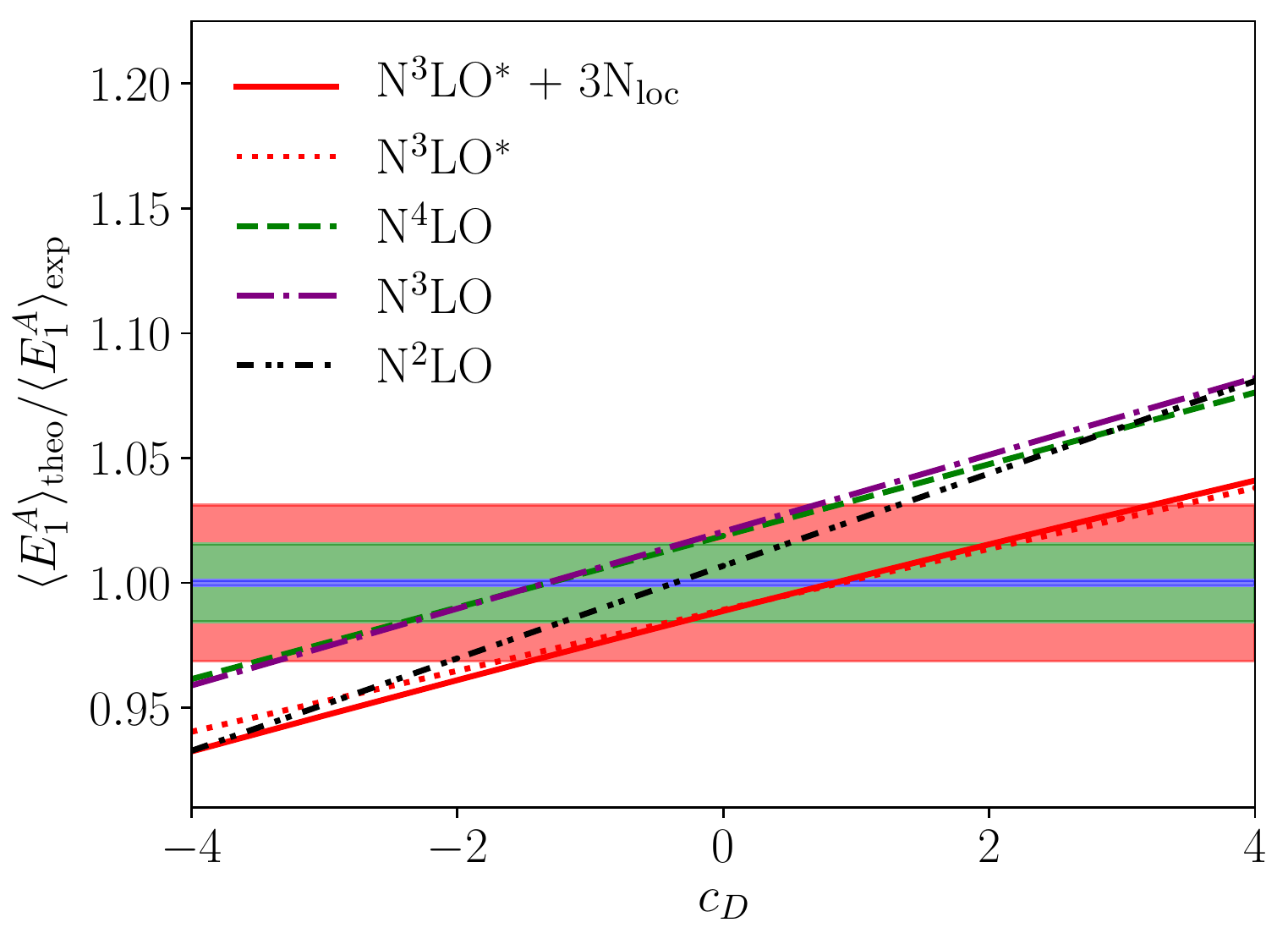}
\caption{Ratio between theoretical and experimental values of $\langle E_1^A\rangle$ for all interaction models used in this work, over a wide range of $c_D$ values. The shaded areas correspond to the estimated uncertainty for the N$^3$LO (red), N$^4$LO (green), and experiment only (blue).}
\label{fig:cDE1curves}
\end{figure}

In Ref.~\cite{Gazit2019} it was shown that the reduced matrix elements of the $J=1$ multipole of the weak axial-vector current $E_1^A$ (entering the description of the $^3$H half-life) present only a weak dependence on the 3N force. Based on this, we first determine the optimal value and confidence interval of $c_{D}$ for which the $E_1^A$ matrix element computed with the bare NN potential reproduces the empirical value $\langle E_1^A\rangle_{\rm exp} = 0.6848 \pm 0.0011$.   The resulting theory-to-experiment ratios as a function of $c_D$, which appears in the two-nucleon contact vertex with an external probe of the nuclear current, are shown in Fig.~\ref{fig:cDE1curves}. 
In a second step, we carry out calculations for the bare NN+3N interaction and use the experimental binding energies of $^3$H and $^3$He as constraints to obtain a functional relationship between $c_D$ and $c_E$, and therefore the optimal value for $c_E$, along with the corresponding confidence interval. 

In the original work of Ref.~\cite{Gazit2019} the uncertainty associated with $\langle E^{A}_1\rangle_{\rm exp}$ was the sole contribution considered in determining the confidence intervals for $c_D$ and $c_E$. Here, we further include the expected uncertainty coming from the truncation of the chiral expansion. Combining the regulator cutoff $\Lambda = 500$ MeV used in this calculation with the corresponding breakdown scale of the chiral expansion $\Lambda_\chi \sim 1$ GeV,
we estimate the chiral expansion parameter as $x=\Lambda/\Lambda_\chi \sim 1/2$ and 
 the expected theoretical uncertainties at N$^3$LO  and N$^4$LO, respectively, as $(1/2)^5\approx 3$\% and $(1/2)^6\approx 1.5$\%~\cite{Entem2003}. 
These are much larger than the uncertainty associated with the empirical value of the $E_A^1$ reduced matrix element and therefore yield a broader range of acceptable values for $c_E$ and (as shown in Fig.~\ref{fig:cDE1curves}) $c_D$. While this estimate of the chiral uncertainties is more rudimentary than  treatments such as those of Refs.~\cite{Epelbaum2015, Furnstahl2015,Melendez2019}, we find it sufficient for the purposes of this work, which only deals with low-energy scattering. In addition, we believe it to be generous enough to account for neglected N$^4$LO contributions to the $E_1^A$ operator~\cite{Baroni2016}. 
The resulting optimal values and confidence intervals for $c_D$ and $c_E$ are summarized in Table~\ref{tab:LECsTable}.

\subsection{NCSMC}
In the NCSMC~\cite{Baroni2013,Baroni2013a}, the microscopic reaction problem is solved by expanding the wave function on 
continuous microscopic-cluster states, describing the relative motion between target and projectile nuclei, and discrete square-integrable states, describing the static composite nuclear system. The idea behind this generalized expansion is to augment the microscopic cluster model, which enables the correct treatment of the wave function in the asymptotic region, with short-range many-body correlations that are present at small separations, mimicking various deformation effects that might take place during the reaction process. The wave function of the system in the partial wave of  spin-parity $J^\pi$  and isospin $T$ is then written as
\begin{align}
\left| \Psi^{J^\pi T} \right\rangle \!\!=\!\! \displaystyle\sum_{\lambda}\! c^{J^\pi T}_\lambda\! \left| \lambda J^\pi T\right\rangle\! +\!\! \displaystyle\sum_{\nu} \int \!\!\!dr \, r^2\, \frac{\gamma^{J^\pi T}_\nu(r)}{r} \mathcal{A}_\nu\left| \Phi_{\nu r}^{J^\pi T}\right\rangle,
\end{align}
where $\lambda$ enumerates the square-integrable energy eigenstates of the aggregate nucleus (here, $^5$He) included in the calculation, $\left| \lambda J^\pi T\right\rangle$, and $\nu$ enumerates the binary-cluster channel states 
of a target (here, $^4\mathrm{He}$) and a projectile (here, a neutron) at a distance $r_{\alpha,n}$, i.e.
\begin{align}
\left| \Phi_{\nu r}^{J^\pi T}\right\rangle  = \left[ 
                                                                    \left( 
                                                                    \left|^4\mathrm{He} \right\rangle 
                                                                    \left|n\right\rangle 
                                                                     \right)^{(s T)} 
                                                                     Y_\ell(\hat{r}_{\alpha,n})
                                                                     \right]^{(J^\pi T)} 
                                                                     \frac{\delta(r-r_{\alpha,n})}{r r_{\alpha,n}}.
\end{align}
Spin-parity and isospin quantum numbers for the target and projectile nuclei are omitted from the notation in the interest of brevity.
The translationally invariant eigenstates of the aggregate, target, and projectile nuclei are all obtained by means of the no-core shell model (NCSM)~\cite{Navratil2000a,Barrett2013} using a basis of many-body harmonic oscillator (HO) wave functions with the same frequency, $\hbar\Omega$, and maximum number of  particle excitations from the lowest Pauli-allowed many-body state, $N_{\rm max}$.
The operator ${\mathcal A}_\nu$ ensures the full antisymmetrization of the microscopic cluster states.
Both the discrete $c_\lambda$ and continuous $\gamma_\nu(r)$ variational amplitudes are unknown, and are obtained by solving a generalized eigenvalue problem in the Hilbert space spanned by the basis states $\left| \lambda J^\pi T\right\rangle$ and $\mathcal{A}_\nu\left| \Phi_{\nu r}^{J^\pi T}\right\rangle$ 
by means of an extension of the microscopic $R$-matrix  method on a Lagrange mesh~\cite{Hesse1998}. While in this section we focused on the formalism for binary reaction processes, the NCSMC can also be applied to the description of three-cluster dynamics~\cite{Romero-Redondo2016,Quaglioni2018}. Additional details on the NCSMC approach can be found in Ref.~\cite{Navratil2016} and references therein.

\subsection{Estimation of the chiral truncation uncertainty}
\label{sec:ChiralDOB}
To estimate the uncertainty due to the truncation of the chiral expansion to a finite order $\kappa$, we adopt the approach described in Ref.~\cite{Melendez2019},  which explicitly accounts for the correlations within continous observables, e.g., cross sections or phase shifts as a function of energy. In this approach,
the value of an observable $O(Q)$ is written as the sum of a term corresponding to the contributions from orders included in the calculation, $O^{(\kappa)}_\mathrm{EFT}(Q)$, and a second term encompassing contributions that are absent due to the truncation of the chiral expansion, $\Delta^{(\kappa)}_\mathrm{EFT}(Q)$, i.e.:
\begin{align}
	O\left(Q\right) 
		&= O^{(\kappa)}_{\rm EFT}\left(Q\right) + \Delta^{(\kappa)}_{\rm EFT}
		\nonumber\\
       	&= O_\mathrm{ref}(Q) \left( \sum_{i=0}^{\kappa} c_i(Q) x^i + \sum_{i=\kappa + 1}^{\infty} c_i(Q) x^i \right)\,. \label{eq:ChiralSum}
\end{align}
Both terms are modeled as a reference scale $O_\mathrm{ref}(Q)$ multiplied by corrections given by the respective chiral expansions 
with respect to the chiral EFT expansion parameter $x$.
At each order $i$ in the summations of Eq.~\eqref{eq:ChiralSum}, the expansion coefficients are assumed to be described by a Gaussian process with a mean of 0 and a kernel 
\begin{equation}
	K(Q, Q') = 
			\sigma_\chi^2 e^{-(Q-Q')^2/2l^2}.
\end{equation}
The hyperparameters $\sigma_\chi$ and $l$ are assumed to be identical for all partial waves.  We fix $\sigma_\chi$ and $l$ by maximimizing the likelihood of the $c_i(Q)$ functions extracted from order-by-order calulations.
We use the mean of the third through fifth orders as the reference scale and set the expansion parameter to be $x=1/2$ as before.  As a result, we find that the hyperparameter $\sigma_\chi$ is of order $1$, indicating that all scales are accounted in our choice of expansion parameter and reference function, and the length scale $l$ is approximately $5\, {\rm MeV}$.

\subsection{Gaussian Stochastic Process Emulator}
\label{sec:GaSP}
In this section we summarize the Gaussian Stochastic Process (GaSP) model 
used to emulate the $n$+$\alpha$ phase shifts and total cross section.  The details of this model are for a functional response such as the $n$-$\alpha$ phase shifts and total cross section investigated in this work, but the same method will be used to fit the scalar $^4$He g.s.\ energy and charge radius as well as the position and width of the $^5$He resonances.  Specifically we implement the method used in Ref.~\cite{gu2018robust} which is robust in terms of the estimation of hyperparameters in the correlation function and uses a Bayesian approach whose details are beyond the scope of this paper.  A GaSP model for a response $y()$ is defined as:
$$ y(\cdot) \sim GaSP( \mu(\cdot), \sigma^2 C(\cdot, \cdot) )$$
with $\mu$ the mean function, $\sigma^2$ the variance and $C$ the correlation function.  Generally, given a set of inputs 
$\mathbf{x}$, the mean function $\mathbf{\mu(x)}$ is given as the expectation value of $y(\mathbf{x})$, and in this case the function is defined over the energy values. This is represented as a linear combination of basis functions $\mathbf{h(x)}$ and regression parameters $\mathbf{\theta}$ so that:
$$ \mathbf{\mu(x)} = E[y(\mathbf{x})] = \mathbf{h(x)\theta}.$$ 
In this work, $y(\mathbf{x})$ is the response (or result) of the theoretical calculation for a specific observable to some inputs $\mathbf{x} =\{c_D,c_E\}$, be it the binding energy or charge radius of $^4$He, or the phase shifts and cross section for \nalpha scattering. 

 This correlation structure is assumed to be stationary, meaning that the correlation between two input locations $\mathbf{x}_1$ and $\mathbf{x}_2$ only depends on the distance $\left| \mathbf{x}_1 - \mathbf{x}_2\right|$ between two observations, and is not based on the location of those observations.   This model is fit using the RobustGaSP R package~\cite{GaSPpackage}.  For scalar responses such as the $^4$He g.s.\ energy the same estimation is applied, however these are no longer functional outputs of the energy.

In the present application, we use a constant mean for $\mu(\mathbf{x})$, but the model first subtracts the average of all predictions. For observables that depend continuously on the energy (phase shifts and cross sections) the average is taken at each energy location.  It is typical when using a GPM to use a simple model for the mean and capture the complexities inherent in the data using the correlation function~\cite{williams2006}.  Here, the correlation function is defined by the Mat\'ern covariance structure. Specifically, we apply a common parameterization referred to as the Mat\'ern $5/2$ so that   
 \begin{align}
C(\mathbf{x_1},\mathbf{x}_2) = & \left(1 + \frac{\sqrt{5}\ | \mathbf{x}_1 - \mathbf{x}_2| }{\gamma} + \frac{5 | \mathbf{x}_1 - \mathbf{x}_2 |^2}{3 \gamma^2} \right)\nonumber\\
&\times  \exp \left( - \frac{ \sqrt{5} | \mathbf{x}_1 - \mathbf{x}_2 | }{ \gamma } \right),
\end{align}
where $\gamma$ is a hyper-parameter that is optimized to obtain a better fit. \\

\section{Application to $n$-$\alpha$ scattering}
\label{sec:results}

\subsection{Parameters of the calculations}
\label{subsec:calcs}
%

%
\begin{figure}[t]
\centering
\includegraphics[width=0.475\textwidth]{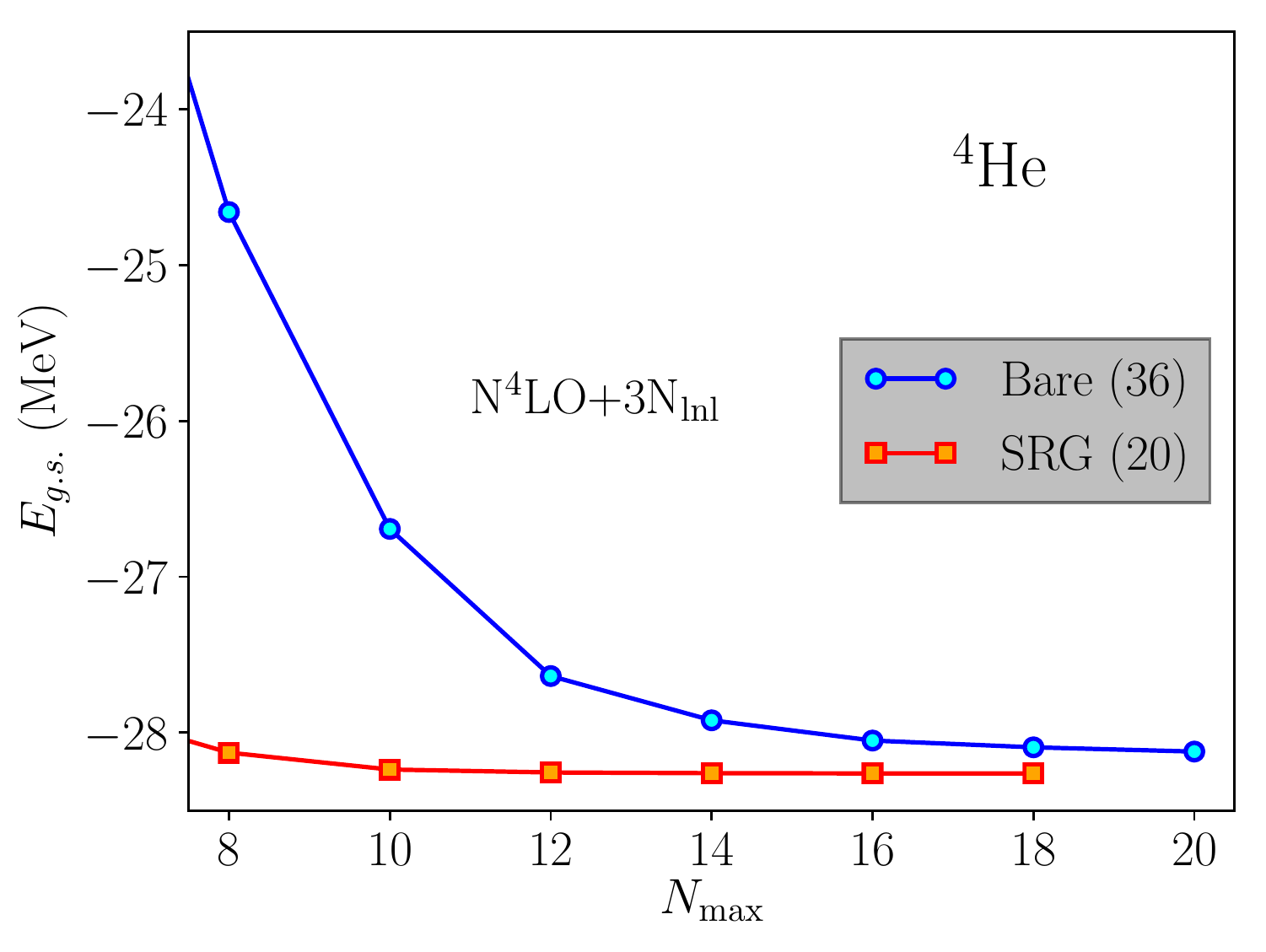}
\caption{\label{fig:helium}Ground-state energy of $^4$He for the N$^4$LO+3N$_{\rm lnl}$ interaction (with $c_D=-1.32$ and $c_E=-0.248$) as a function of the HO model space size $N_{\rm max}$ obtained using the bare Hamiltonian and HO frequency $\hbar\Omega=36$ MeV (circles), and the SRG-evolved Hamiltonian at the resolution scale $\lambda_{\rm SRG}=2.0$ fm$^{-1}$ and $\hbar\Omega=20$ MeV (squares). Both curves are the results of calculations on a four-body translational invariant HO basis.}
\label{fig:n4loHe4}
\end{figure}
In this section we provide the details of the NCSMC calculations used to construct the GPMs of $n$-$\alpha$ scattering for the N$^4$LO+3N$_{\rm lnl}$ and  N$^3$LO$^*$+3N$_{\rm loc}$ chiral interaction models.


To accelerate the convergence of the NCSMC calculation, for each of the design 
runs we soften the NN+3N Hamiltonian by performing a similarity renormalization group (SRG)~\cite{Jurgenson2011,Jurgenson2013} transformation in three-nucleon space. 
We adopt the resolution scale $\lambda_{\rm SRG}=2$ fm$^{-1}$, for which the effect of induced four- and higher-body forces has been shown to be negligible in light nuclei for the N$^3$LO$^*$+3N$_{\rm loc}$ interaction. 
This value for the resolution scale was also adopted in several NCSMC calculations of light-nuclei scattering and reactions, and is therefore useful to estimate the robustness of previously obtained results.  As a test, we computed the binding energy of $^4$He for the N$^2$LO+3N$_{\rm lnl}$, N$^3$LO+3N$_{\rm lnl}$ and N$^4$LO+3N$_{\rm lnl}$ chiral models with both the bare and SRG-evolved Hamiltonian and found that the contribution of four-nucleon induced forces is on the order of 100 keV. The convergence of the two calculations with respect to the size of the HO model space is shown in Fig.~\ref{fig:helium}. We note that with the SRG-evolved Hamiltonian the g.s.\ energy is  already converged at $N_{\rm max}=10$.

NCSMC calculations of $n$-$\alpha$ scattering require as input the NCSM eigenstates of the $^4$He target and of the composite $^5$He system. In this work we used up to the lowest ten states in each of the negative and positive parity sectors of the $A=5$ system, covering an excitation energy range of up to 25 MeV.  Considering the negligible effect of target excited states on the proton-$\alpha$ scattering phase shifts of Ref.~\cite{Hupin2014}, for the present analysis of uncertainties we only include the ground state of the $^4$He nucleus.
 Further, all design runs are performed starting from the above-mentioned SRG-evolved Hamiltonians in an HO model space with $\hbar\Omega=20$ MeV and  $N_{\rm max}=10 (11)$ for the $^4$He ground state and $^5$He negative-parity states ($^5$He positive-parity states and $n$-$\alpha$ relative motion). Previous calculations based on the N$^3$LO$^*$+3N$_{\rm loc}$ interaction indicate that this choice for the wave functions yields adequate convergence for the present study. Finally, the description of the $n$-$\alpha$ reaction cross section for neutron incident energies above 20 MeV, which would require the inclusion of deuterium-$^3$H channels in the NCSMC calculation~\cite{Hupin2019}, lies outside the scope of this work.

\subsection{Convergence of the chiral expansion}%
\begin{table*}[t]
\begin{center}
\begin {ruledtabular}
\caption{\label{tab:LECsBErch}Calculated $^4$He g.s.\ energy (in MeV), point-proton radius (in fm) and charge radius (in fm), obtained using the chiral interaction models of Table~\ref{tab:LECsTable} with the  optimal values for $c_D$ and $c_E$, compared to experiment. Chiral uncertainties obtained in the naive approach described in Sec.~\ref{subsec:interactions} are shown after the $\pm$ sign, whereas estimates of the model errors are in parentheses. For the charge radius the error in parenthesis also includes the uncertainties from $R_{n}$ and $R_{p}$ (see text for additional details). 
}
\begin{tabular}{ l  c c c c c }
Chiral Model& $c_D^\mathrm{opt}$ & $c_E^\mathrm{opt}$ &  $E_{\rm g.s.}$($^4$He) & $r_{\rm pp}$($^4$He)& $r_{\rm ch}$($^4$He) \\[2mm]
\hline\\[-3mm]
LO               &      -   &    -      &   ~\;\!$-40.0873(6)\pm10.02$ & 1.0869(1) &$1.344(2)\pm0.22$ \\[2mm] 
NLO                                  &      -   &    -      &  ~$-27.542(1)\pm3.44$   & 1.475(1)&$1.674(2) \pm0.171$\\[2mm] 
N$^2$LO+3N$_{\rm lnl}$ & $-0.37$ & $-0.189$ & ~$-27.827(5)\pm1.74$   & 1.457(2)&$1.658(2)\pm0.085$\\[2mm]
N$^3$LO+3N$_{\rm lnl}$ & $-1.33$ & $-0.413$ & ~$-27.90(2)\pm0.87$   & 1.472(2)&$1.671(2)\pm0.043$\\[2mm]
N$^4$LO+3N$_{\rm lnl}$ & $-1.32$ & $-0.248$ & ~$-28.14(2)\pm0.44$   & 1.468(2)&$1.668(2)\pm0.021$\\[2mm]
N$^4$LO+3N$^b_{\rm lnl}$ & $-1.32$ & $-0.627$ & ~$ -27.84(2)\pm0.44$   & 1.492(2)&$1.689(2)\pm0.021$\\[2mm]
\hline
Expt.                                 &      -   &    -      &  \quad$-28.296$\phantom{()} & &\quad 1.6755 (28) \,\,\, 
\end{tabular}
\end{ruledtabular}
\end{center}
\end{table*}
We begin by discussing the order-by-order convergence of the binding energy and charge radius of the $^4$He ground state for the interactions of Ref.~\cite{Entem2017}, combined (from next-to-next-to leading order and above) with the 3N force at N$^2$LO using the optimal $c_D$ and $c_E$ values presented in Table~\ref{tab:LECsTable}. 
The results, obtained starting from the bare Hamiltonian and working in a translational-invariant four-body HO basis (including up to $N_{\rm max} = 20$ excitations above the g.s.\ configuration), are summarized in Table~\ref{tab:LECsBErch} where we also report the model uncertainties (in parentheses) followed by
a naive estimate (see Sec.~\ref{subsec:interactions}) of the uncertainties arising from the truncation of the chiral expansion 
(shown with the $\pm$ sign). 
Here, we refer to model uncertainties as any remaining uncertainty related to the convergence of the calculation with respect to the model parameters.
We note that the approach for estimating the chiral truncation uncertainty described in Ref.~\cite{Epelbaum2015} yields  an almost identical outcome, provided the LO results are not taken into account in the estimate. 
 At next-to-leading order, the $^4$He is under-bound by about 2 MeV. 
 At each increasing order, the system becomes progressively more bound, yielding a binding energy of 28.14(2) MeV at N$^4$LO. 
 The $\sim200$ keV discrepancy with respect to  the experimental value is well within the 1.5\% uncertainty naively expected at this order, shown in Table~\ref{tab:LECsBErch}.
 The chiral uncertainty dominates over the model uncertainty in the predictions for the g.s.\ energy. 
 The charge radius $r_\mathrm{ch}$ was obtained from the point-proton radius $r_{\rm pp}$ (also shown in Table~\ref{tab:LECsBErch}) as $r_\mathrm{ch}^2=r_\mathrm{pp}^2+R_n^2+R_p^2+\frac{3\hbar^2}{4m^2_pc^2}$~\cite{EkJa15}, with $\frac{3\hbar^2}{4m^2_pc^2}=0.033$, $R^2_n=-0.1149(27)$~\cite{Angeli2013}, and $R_p = 0.8414(19)$~\cite{Tiesinga2020}.
 For the charge radius we find a rapid convergence, with the NLO result already close to the empirical value.
 At all orders the chiral uncertainty again dominates, with the experimental value being within the chiral error bands in all cases.
 
\begin{figure}[b]
\centering
\includegraphics[width=0.475\textwidth]{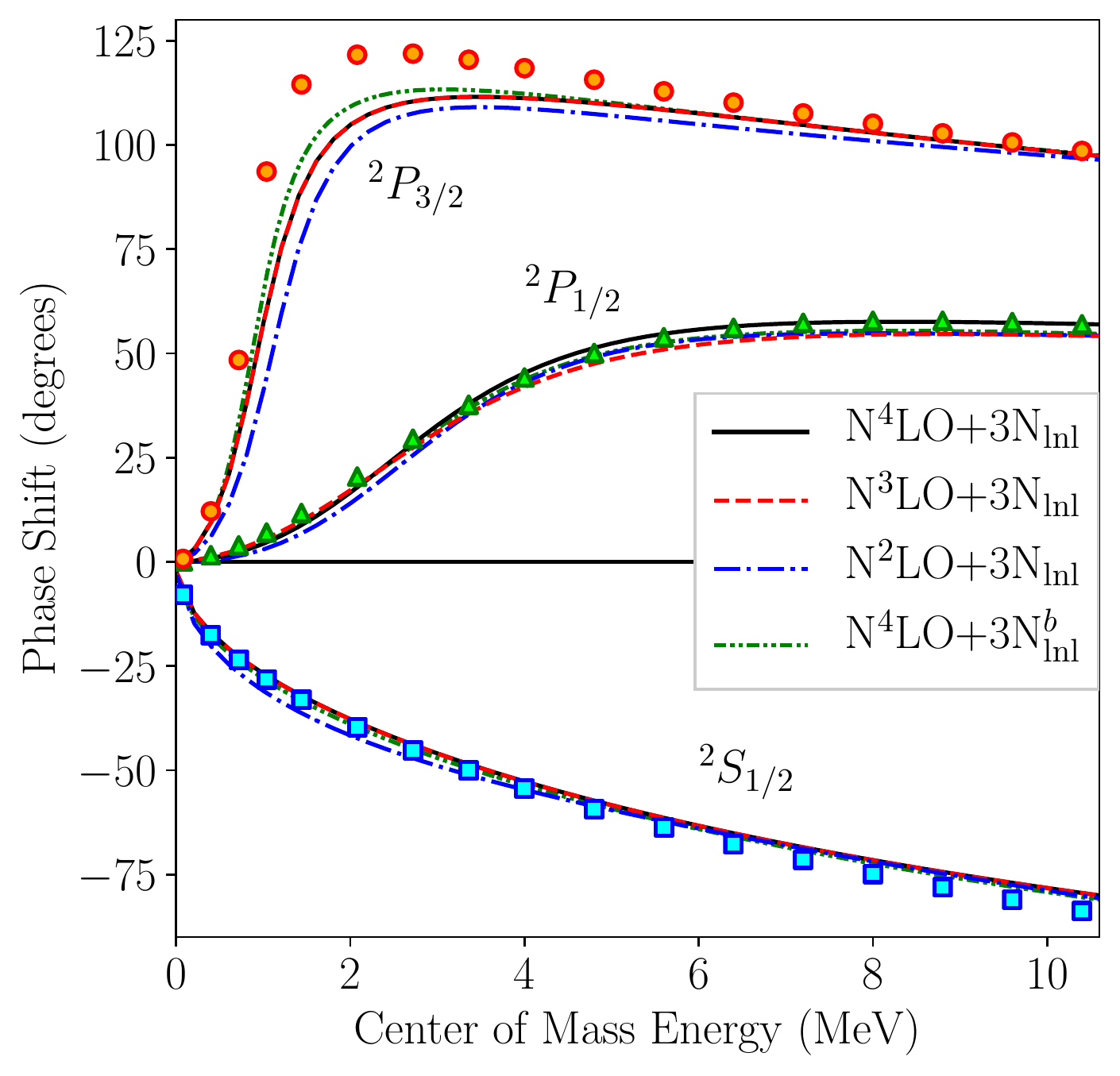}
\caption{Evolution  of the $n$-$\alpha$ phase shifts (lines) from third to fifth order of the chiral expansion compared to the empirical phase shifts obtained from an accurate $R$-matrix analysis of $A=5$ reaction data~\cite{Hale} (symbols).}
\label{fig:ChiralComparison}
\end{figure}
\begin{figure}[b]
\centering
\includegraphics[width=0.475\textwidth]{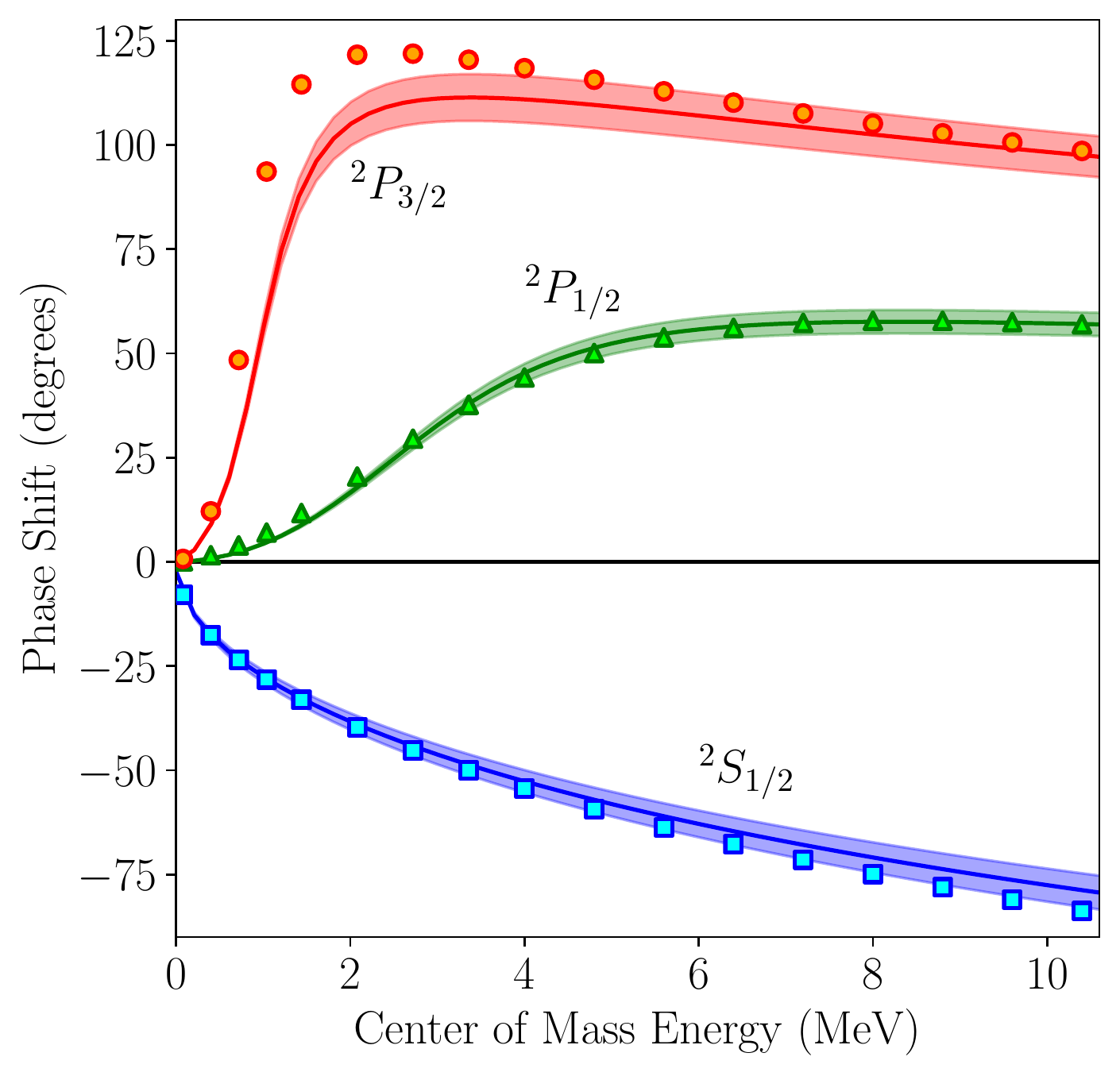}
\caption{Bayesian estimation of the uncertianty induced by the truncation of the chiral expasnsion.  The bands correpsond to a 90\% degree of belief interval estastimate at the fifth order in the chiral expansion. 
\label{fig:ChiralDOB}
}
\end{figure}
The evolution of the $n$-$\alpha$ phase shifts starting from the third order of the expansion is shown in Fig~\ref{fig:ChiralComparison}.  The N$^3$LO+3N$_{\rm lnl}$ and N$^4$LO+3N$_{\rm lnl}$ results for the $J^\pi=3/2^-$ ($^2P_{3/2}$) resonance and  $1/2^+$ ($^2S_{1/2}$) repulsive phase shifts are indistinguishable from each other, while they deviate slightly for the $1/2^-$ ($^2P_{1/2}$) resonance. In all shown partial waves, the convergence is fairly rapid, yielding a reasonably good description of the empirical phase shifts from an accurate $R$-matrix analysis of $A=5$ reaction data~\cite{Hale} at low energies in all but the $^2P_{3/2}$ ground-state resonance of $^5$He. 
Figure~\ref{fig:ChiralDOB}, showing the results of the Bayesian estimate of the chiral uncertainty discussed in Sec.~\ref{sec:ChiralDOB}, suggests that such disagreement between empirical phase shifts and the NCSMC predictions using the N$^4$LO+3N$_{\rm lnl}$ interaction cannot be attributed to chiral truncation uncertainties. 
While in the present parameterization of the N$^2$LO 3N force we adopt the effective $\pi N$ LECs recommended in Ref.~\cite{Entem2017} in an effort to account for missing higher-order terms, we cannot exclude that the inconsistency between the chiral expansion for the NN and 3N forces may be at the root of such disagreement. 
In particular, the insufficient splitting between the $^2P_{3/2}$ and $^2P_{1/2}$ partial waves points to the need for additional spin-orbit strength in the 3N force, which could be supplied by the N$^4$LO  3N contact interactions derived by Girlanda et al.~\cite{Girlanda2011}. At the same time, such corrections would be unexpectedly large for this order. To investigate this somewhat surprising result further, we considered the additional chiral Hamiltonian model obtained by combining the N$^4$LO NN potential and the N$^2$LO 3N force with the (bare) $\pi N$ LECs consistent with the NN sector, dentoted N$^4$LO+3N$^b_\mathrm{lnl}$. The optimal values for $c_D$ ($-1.32$) and $c_E$ ($-0.627$) were obtained following the procedure described in Sec.~\ref{subsec:interactions}.
With this interaction, we find the $^4$He nucleus to be underbound with respect to experiment by about 0.5 MeV (see Table~\ref{tab:LECsBErch}), which is somewhat larger than the naive chiral truncation uncertainty at N$^4$LO. However, given that in this case no effort is being made to account for higher-order terms of the 3N force, we may be underestimating the chiral uncertainty. The corresponding results for the $n$+$\alpha$ phase shifts, shown in Fig.~\ref{fig:ChiralComparison},  seem to indicate a somewhat larger spin-orbit strength though still insufficient to reproduce the empirical $^2P_{3/2}$ resonance.
In the following section, we will show that the observed discrepancy with the phenomenological phase shifts cannot be attributed to a poor determination of the LECs $c_D$ and $c_E$. 

\subsection{Sensitivity to $c_D$ and $c_E$}
\begin{figure}[t]
\centering
\includegraphics[width=0.475\textwidth]{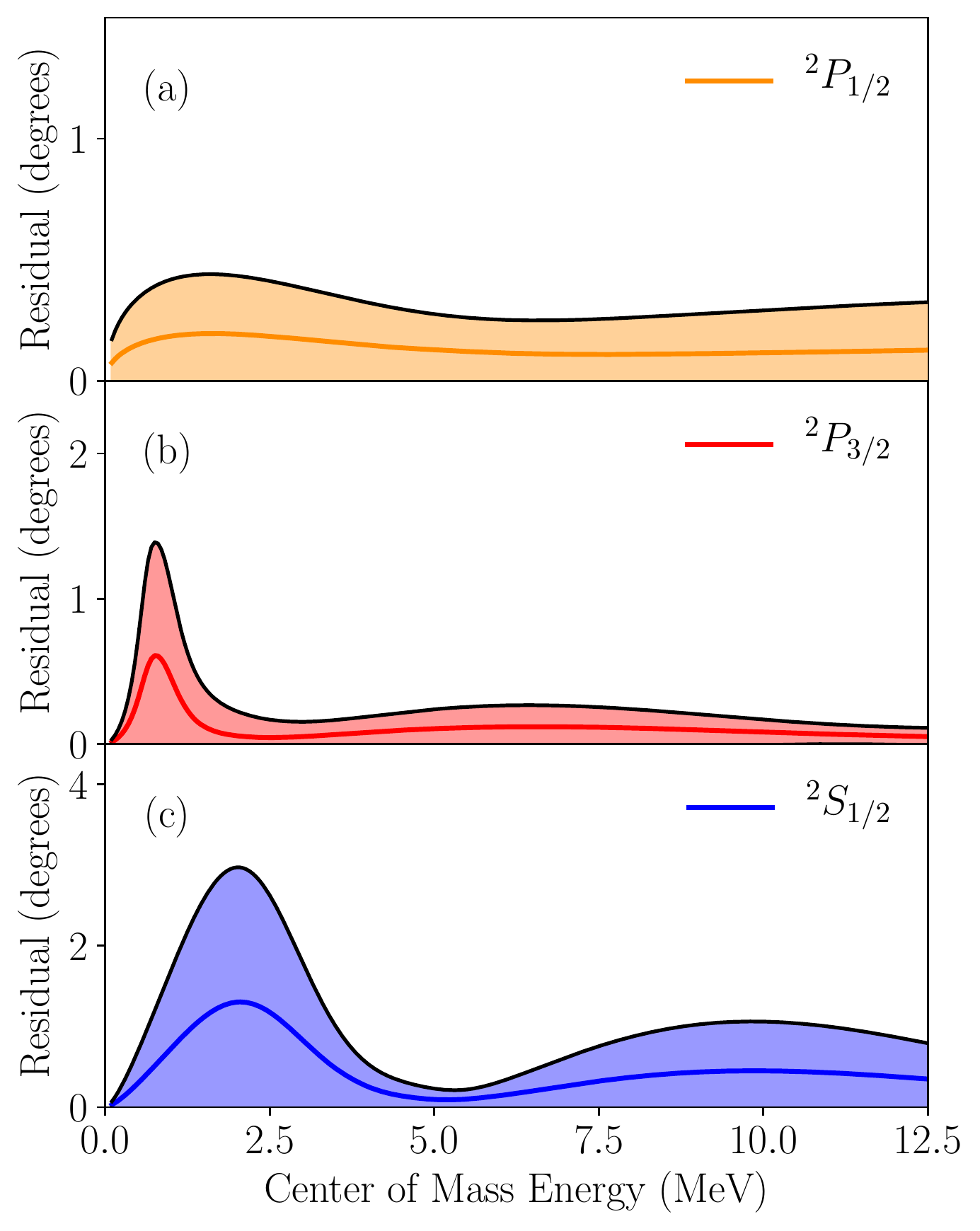}
\caption{Mean and standard deviation of the residual between the GaSP emulators obtained using all but one design points  and the NCSMC calculation for the omitted design point. Panels (a), (b) and (c) show results for the $^2P_{1/2}$, $^2P_{3/2}$ and $^2S_{1/2}$  phase shift, respectively, obtained with the N$^3$LO$^*$+3N$_\mathrm{loc}$.}
\label{fig:Residual}
\end{figure}
\begin{figure}[b]
\centering
\includegraphics[width=0.475\textwidth]{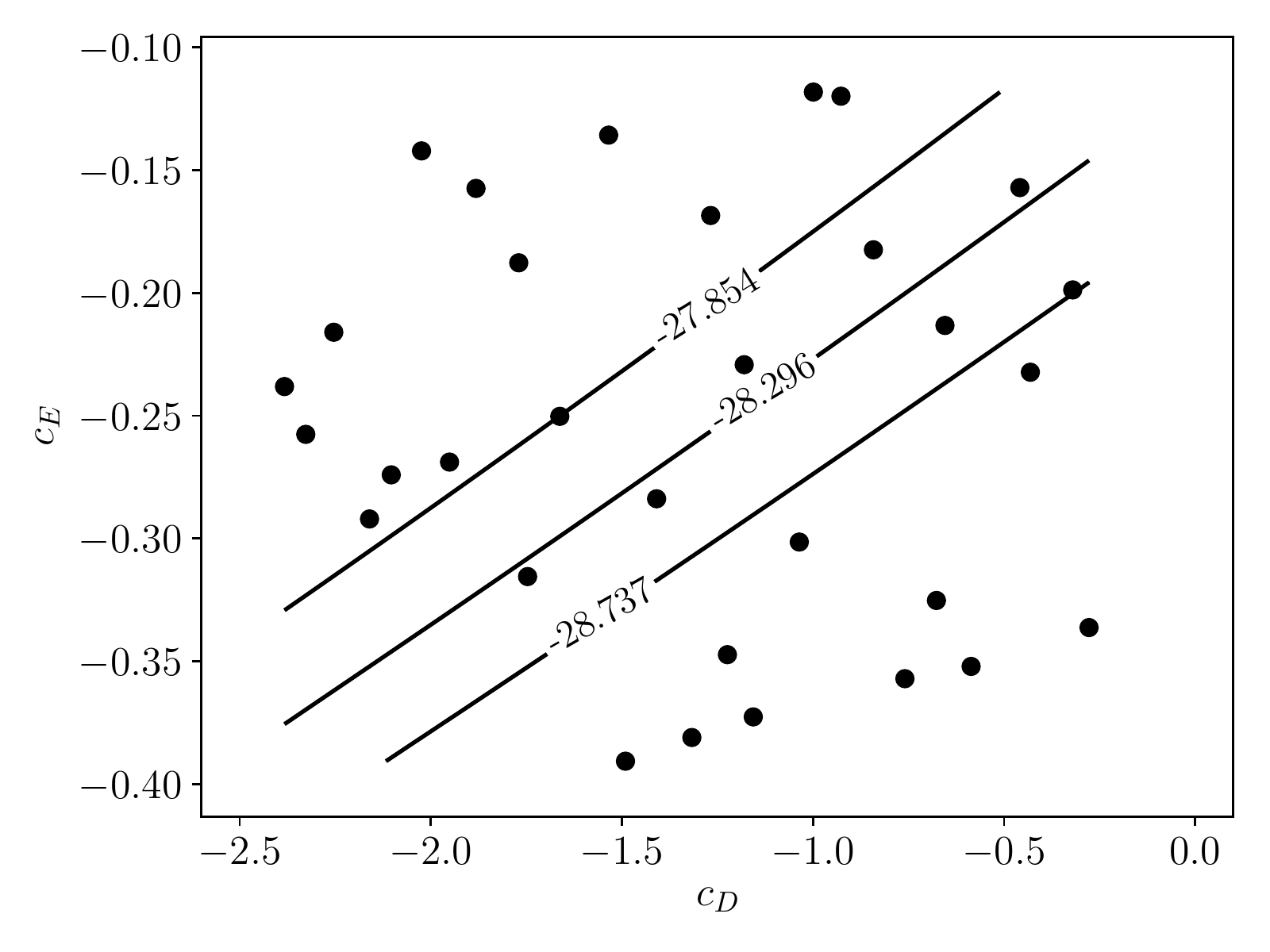}
\caption{Scatter plot of the design points used in the N$^4$LO+3N$_\mathrm{lnl}$ GPM, along with contours for the $^4$He binding energy and the $\pm 1.5\%$ interval from the chiral order uncertainty. }
\label{fig:n4loHe4}
\end{figure}
\begin{figure}[t]
\includegraphics[width=0.475\textwidth]{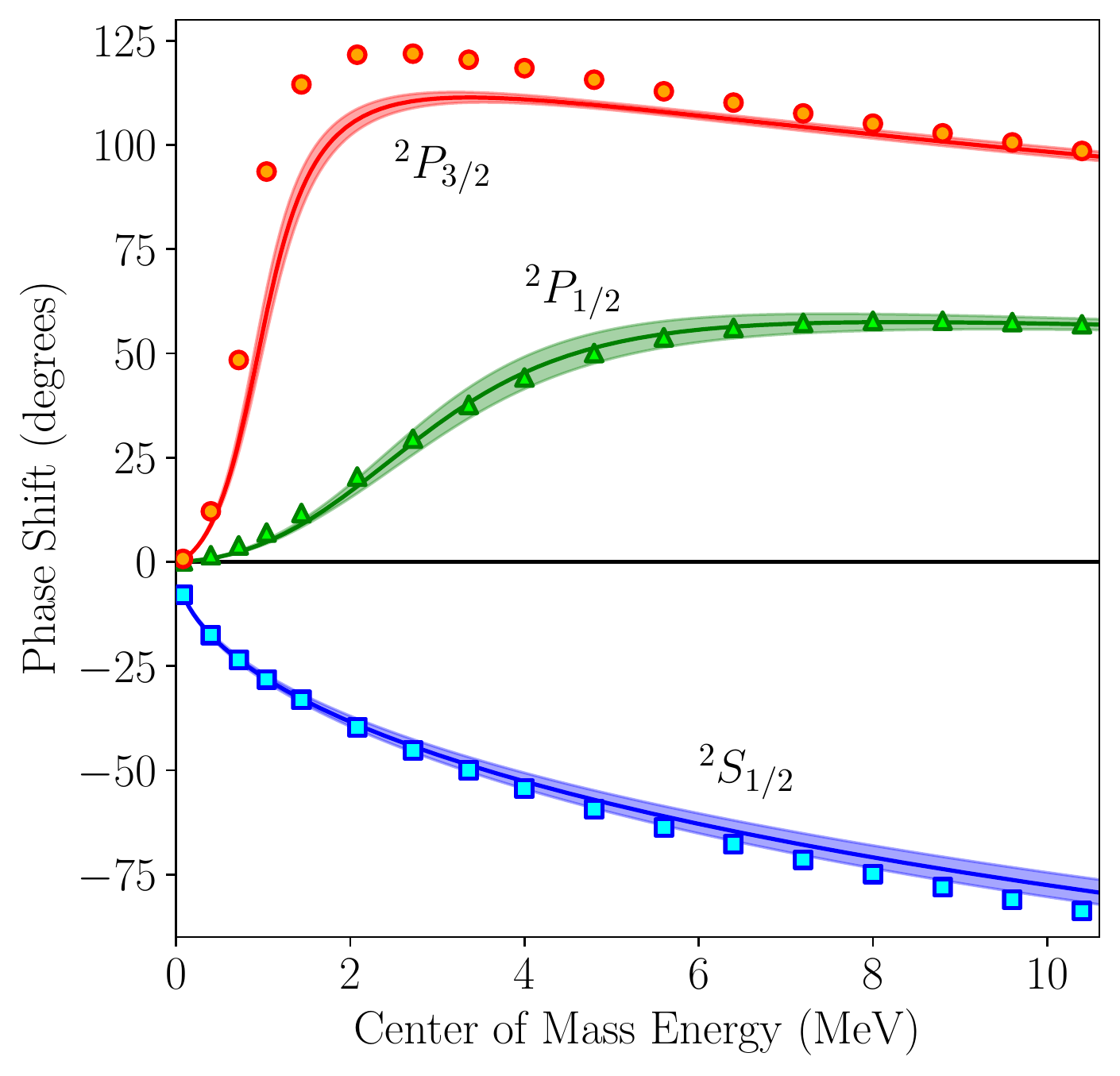}
\caption{Predicted phase shifts (solid lines) from the GPM for the N$^4$LO+3N$_\mathrm{lnl}$ interaction, along with the 95\% confidence intervals (shaded) after sampling the full range of acceptable LECs.}
\label{fig:n4loPhaseShifts}
\end{figure}
\begin{figure}[b]
\centering
\includegraphics[width=0.475\textwidth]{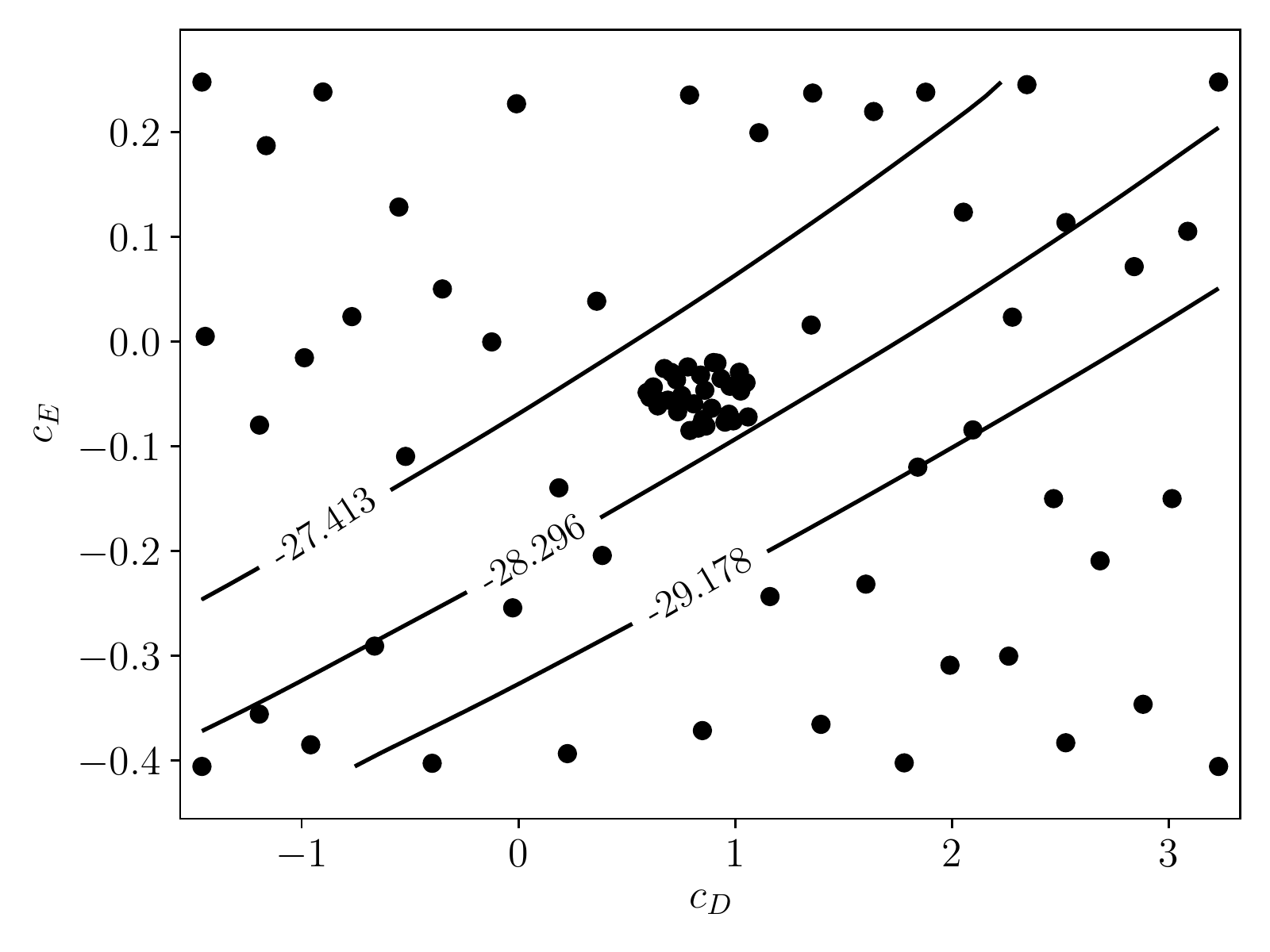}
\caption{Same as Fig.~\ref{fig:n4loHe4}, but for the N$^3$LO$^*$+3N$_\mathrm{loc}$ interaction. The contours now span the $\pm$3\% range consistent with this chiral order.}
\label{fig:n3loHe4}
\end{figure}

We now discuss the sensitivity of the computed $^4$He g.s.\ properties and $n$-$\alpha$ phase shifts to the variation of the $c_D$ and $c_E$ 3N-force LECs within the uncertainty bands determined in Sec.~\ref{subsec:interactions}. We perform such study for both the N$^4$LO+3N$_{\rm lnl}$ and N$^3$LO$^*$+3N$_{\rm loc}$ chiral models.

Within the range of $(c_D, c_E)$ values shown in Table~\ref{tab:LECsTable}, we assume a uniform prior distribution and sample 31 and 81 points in a Latin Hypercube design \cite{santner2003} for the N$^4$LO+3N$_{\rm lnl}$  and N$^3$LO$^*$+3N$_{\rm loc}$, respectively. As an example, a scatter diagram of the design points for the N$^4$LO+3N$_{\rm lnl}$ chiral model is shown in Fig.~\ref{fig:n4loHe4}. 
Calculations of the $^4$He g.s.\ energy, charge radius, $n$-$\alpha$ scattering phase shifts (see Sec.~\ref{subsec:calcs}) and total cross section for each of the NN+3N Hamiltonians corresponding to the design points are then used to construct corresponding GPMs using the GaSP model described in Sec.~\ref{sec:GaSP}.
 To test the accuracy of the GaSP model, we constructed emulators using all but one of the design runs and compared the resulting prediction with the omitted data. An example of the mean residual after iterating over all design points is shown in Fig.~\ref{fig:Residual} for the $n+\alpha$ phase shifts computed with the N$^3$LO$^*$+3N$_{\rm loc}$ interaction. 
The largest residuals occur in regions where the phase shifts change quickly and are largest when modeling the $^2S_{1/2}$ channel. 
Above 5 MeV, this is likely due to the sensitivity of the phase shift in this channel on a higher-lying resonance which is not properly described in the NCSMC calculations presented here. 
A sharp resonance appears at $\sim 20$ MeV in correspondence of the eigenenergy of the $^5$He state dominated by deuterium-$^3$H clustering. This is due to the missing escape width for the deuterium-$^3$H channel, which is not included in the present calculation. The position of this resonance varies significantly for each pair of LECs, causing the $s$-wave phase shifts to demonstrate a large variance at higher energies. We note that the explicit inclusion of the deuterium-$^3$H channel is also required to  accurately reproduce the position of the resonance.
At lower energy, the $^2S_{1/2}$ phase shift is affected by the existence of a Pauli-blocked state, leading to the observed somewhat increased difficulty of the GaSP to describe it.

Figure~\ref{fig:n4loHe4} shows the design points distributed in the $(c_D, c_E)$ plane along with contours extracted from the GPM both for the exact binding energy of $^4$He, as well as the estimated $\pm1.5\%$ uncertainty band for this order.
Any pair of LECs that falls within the band predicts the binding energy of $^4$He within the uncertainty limits predetermined for the theory. Because of this, this $A=4$ observable alone does not provide a unique constraint for the two LECs.  
\begin{figure}
\centering
\includegraphics[width=0.475\textwidth]{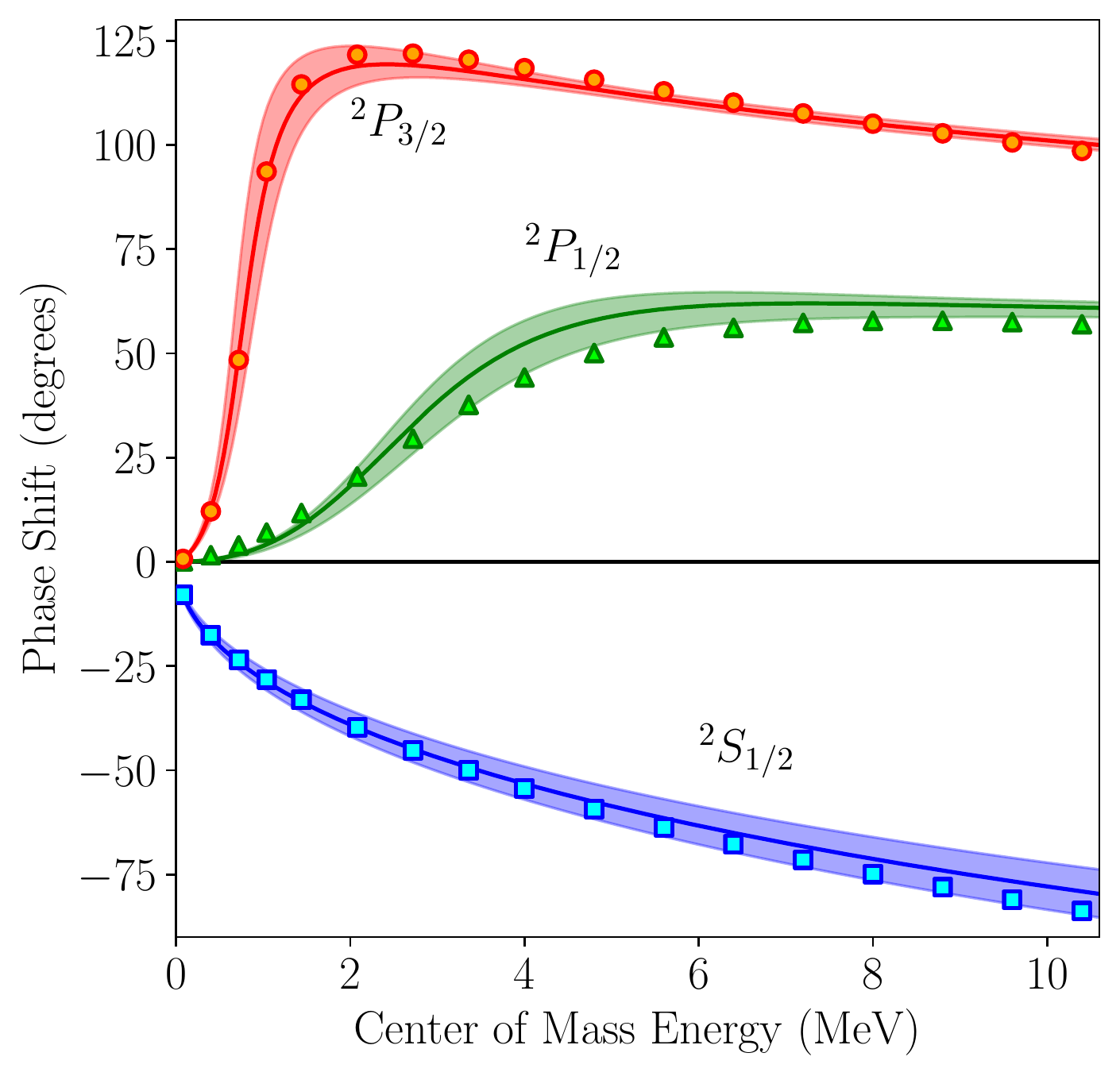}
\caption{Same as Fig.~\ref{fig:n4loPhaseShifts}, but for the N$^3$LO$^*$+3N$_\mathrm{loc}$ interaction. }
\label{fig:n3loPhaseShifts}
\end{figure}

Drawing samples from a uniform distribution in the allowed $(c_D,c_E)$ plane and using the GaSP model described in Sec.~\ref{sec:GaSP} to predict the phase shifts at each sample point we extract the mean prediction for the phase shifts as well as the 95\% confidence intervals; the results are plotted in Fig.~\ref{fig:n4loPhaseShifts}.
The low-energy $J^\pi=3/2^-$ phase shifts are not well described within the confidence interval, pointing to either a deficiency in the two-body sector of the interaction or an underestimation of the chiral error which would lead to larger limits for the LECs. The latter can be understood as the result of mixing orders in the two- and three-body sectors of the interaction. Nevertheless, we have established from the previous section that the order-by-order convergence seems good, and therefore it does not seem as such an increase in the limits would be justified. 

For the N$^3$LO$^*$+3N$_\mathrm{loc}$ interaction we adopt a slightly modified strategy in selecting the design points for the GPM.  In order to achieve lower statistical uncertainties for the GPM predictions, we sample 31 points in the region of the $(c_D, c_E)$ plane that would be determined if one were to omit the chiral order uncertainty in the $\beta$-decay estimate of $c_D$.  We then supplement another 50 design points, using a simulated annealing algorithm \cite{kirkpatrick1983} to optimize the minimax space filling criterion \cite{johnson1990}, to cover the entire plane. The resulting set of design points is shown in Fig~\ref{fig:n3loHe4}, along with the extracted contours for the g.s.\ energy of $^4$He.  
The large increase in confidence intervals between the first 31 and the additional 50 points demonstrates the significance of including the chiral order uncertainty in the fitting process for the LECs, allowing for a broader range of predicted $^4$He energies, that now include the correct experimental value. 
\begin{figure*}
\centering
\includegraphics[width=1.\textwidth]{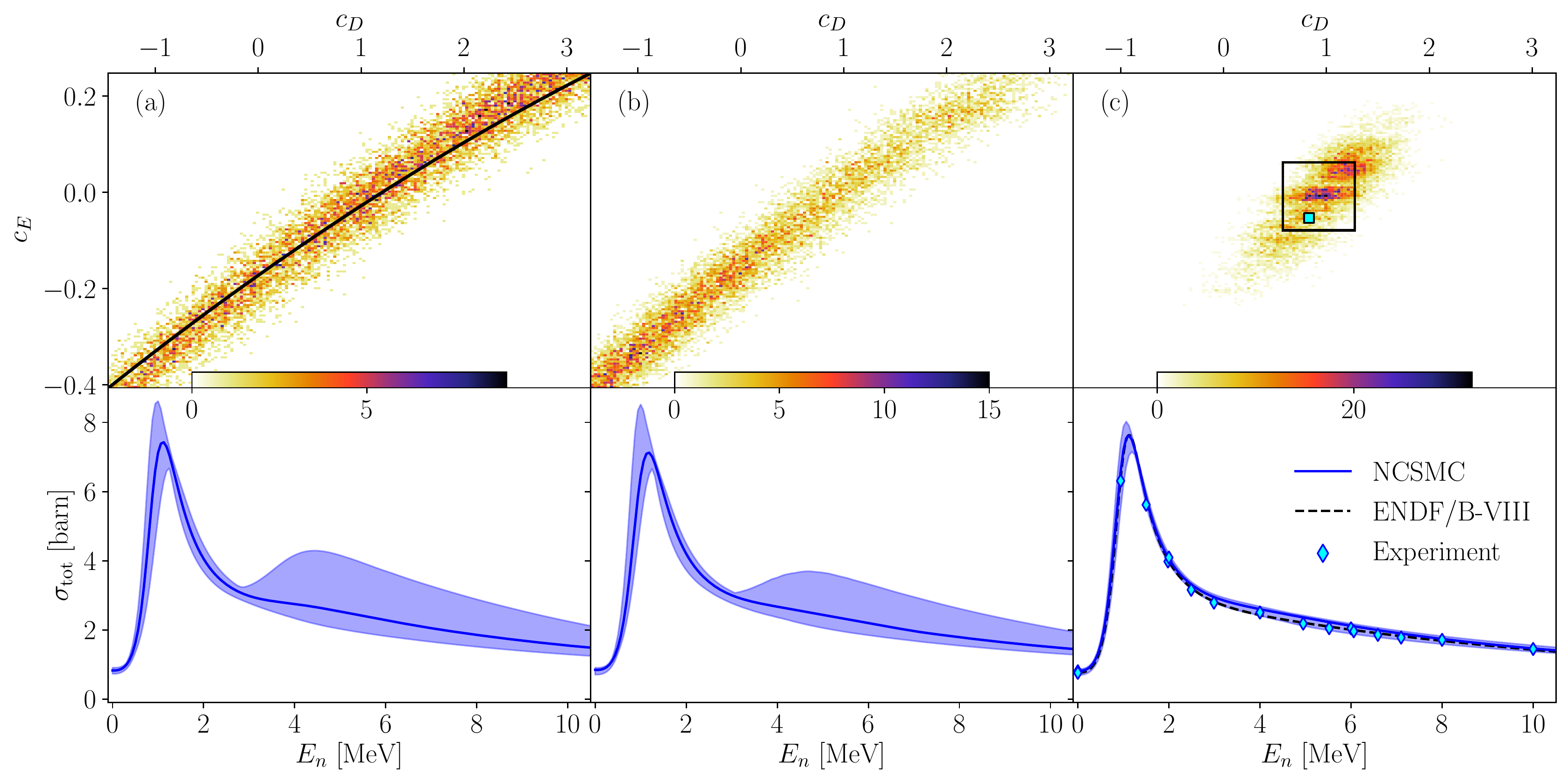}
\caption{\label{fig:posteriors} Posterior distributions of $c_D$ and $c_E$ values after incrementaly using the binding energies of $^3$H and $^4$He, the charge radius of $^4$He, and finally, positions and widths of the $J^\pi=3/2^-,1/2^-$ \nalpha resonances, as constraints. For each case we also show the resulting prediction for the cross section along with its 95\% confidence interval. The black line in panel (a) is the trajectory of $c_D$-$c_E$ values that reproduce on average the $^3$H and $^3$He binding energies. In panel (c), the cyan square is the location of the previously determined optimal $c_D$, $c_E$ values~\cite{Gazit2019}, and the rectangle gives the maximum likelihood a posteriori LECs (see text).}
\end{figure*}
Overall, we observe a similar trend as in Fig.~\ref{fig:n4loHe4}, with the $(c_D,c_E)$ pairs reproducing the empirical g.s.\ energy (within the $3\%$ uncertainty band for this order) lying along a diagonal band. However, in this case the allowed values for the LECs are both positive and negative, leaving the character of the corresponding 3N-force diagrams (attractive/repulsive) undetermined. This ambiguity is purely due to the increased uncertainty limits brought about by including the chiral order uncertainty in determining the $\beta$-decay half-life and would not necessarily exist at higher orders.

The corresponding $n$-$\alpha$ phase shifts for the \ntlos+3N$_\mathrm{loc}$ are shown in Fig.~\ref{fig:n3loPhaseShifts}, with the empirical data lying within the uncertainty estimates of the chiral prediction. This set of calculations appears to have an overall better agreement with the empirical data than 
those performed with the N$^4$LO+3N$_\mathrm{lnl}$ interaction. Since the 3N sector of the interaction is identical and is varied within uncertainties in both cases, a remaining possible source of discrepancy is the different parametrization chosen for the NN LECs. 
The uncertainty band is broader than in Fig.~\ref{fig:n4loPhaseShifts}, reflecting the doubling of the chiral uncertainty in the determination of $c_D$ because of the lower chiral order. 
The broadening of the  uncertainty band for the $J^\pi=1/2^+$ phase shift at higher energy is an artifact due to a higher-lying resonance in this channel which is not properly described in the present choice for the NCSMC model space, as discussed earlier in this section.

\subsection{Posterior distributions for $c_D$ and $c_E$}

As a last step, starting from the design points for the N$^3$LO$^*$+3N$_\mathrm{loc}$ interaction, we construct GPMs for the $^3$H and $^4$He binding energies, the $^4$He charge radius, as well as for the positions and widths of the $J^\pi = 3/2^-,1/2^-$ $P$-wave resonances obtained from the scattering calculation using the same method as in Ref.~\cite{DohetEraly2016}. 
We then use these emulators, along with the experimentally observed values for these quantities, to obtain a posterior distribution for the LECs. 

 We define a normal log-likelihood function 
\begin{align}
\chi^2 = -\frac{1}{2}\displaystyle\sum_i \frac{(O_i^{th}-O_i^{\rm exp})^2}{\sigma_{\rm th}^2 + \sigma_{\rm exp}^2},
\label{eq:chisquare}
\end{align}
that treats the theoretical ($\sigma_{\rm th}$) and experimental ($\sigma_{\rm exp}$) uncertainties as uncorrelated, but both equally contributing to the overall uncertainty of an observable $O_i$~\cite{Brown2006}. The theoretical uncertainty is taken as an uncorrelated sum of the uncertainty coming from the GPM prediction as well as the expected chiral uncertainty for each observable. For the resonance information, we further take into account the tolerance limit of the minimization-based method used for extracting them~\cite{DohetEraly2016}. 
In the overall estimate, we do not consider any uncertainty coming from model uncertainties 
in the method used to solve the many-body scattering problem as the calculations shown here are converged.  The theoretical uncertainty is substantially larger than the experimental uncertainty contributions in all cases allowing us to ignore the latter in the sum. The observables considered here, along with their estimated uncertainties are listed in Table~\ref{tab:FitParameters}. 
\begin{table}[t]
\begin{center}
\begin {ruledtabular}
\caption{\label{tab:FitParameters} Observables used in the log-likelihood function, along with their estimated uncertainties. The column labeled Theory corresponds to the value for each parameter as sampled from the respective GPM with the probability distribution function shown in Fig.~\ref{fig:posteriors}, panel (c).}
\begin{tabular}{ l  c c c c }
&	Expt.	& $\sigma_{\rm exp}$	&	$\sigma_{\rm th}$	& Theory \\ [2mm]
\hline
Tritium\\
$E_{\rm g.s.}$ (MeV)          		&  $-8.482$     &  $\sim$0.001      &  0.085 & $-8.46(4)$ \\[2mm]
Helium-4\\
$E_{\rm g.s.}$ (MeV)       		&  $-28.295$   &  $\sim$0.001     & 0.280 &  $-28.47(24)$ \\
$r_{\rm ch}$ (fm)   		& 1.6755   	&  0.0028      & 0.0167 &  $1.707(5)$ \\[2mm]
Helium-5\\
$E_{\rm R}^{3/2^-}$ (MeV) 		& 0.798 	& - & 0.025 &  $0.756(13)$ \\
$\Gamma_{\rm R}^{3/2^-}$ (MeV)  	& 0.648  	& - & 0.025 &  $0.663(10)$ \\
$E_{\rm R}^{1/2^-}$ (MeV)		& 2.068  	& - & 0.167 &  $2.32(13)$ \\
$\Gamma_{\rm R}^{1/2^-}$ (MeV) 	& 3.18  	& - & 0.250  &  $3.84(20)$ \\
\end{tabular}
\end{ruledtabular}
\end{center}
\end{table}

The log-likelihood function is evaluated using Markov chain Monte Carlo (MCMC) samples to obtain the posterior distribution. We select a uniform prior over the chosen range of parameter values. 
To obtain the samples we implement a delayed rejection adaptive Metropolis (DRAM) MCMC algorithm~\cite{haario2006dram}.  This updates the covariance of the proposal distribution during the burn-in period and increases the efficiency of the sampling during MCMC.  We generate $4\times10^6$ samples and discard the first $10^4$ as burn-in.  The chain is thinned so that only $10^4$ samples are kept.  This was done to ensure that the resulting posterior samples are uncorrelated. 

To highlight the effects of each different observable on the posterior distribution for the LECs we progressively extend the summation in Eq.~\eqref{eq:chisquare} to include, the $^3$H g.s.\ energy and first only the g.s.\ energy of $^4$He, then also the charge radius of $^4$He, and finally the $^5$He resonance positions and widths. 
The posterior distribution at each step can be seen in the top row panels of Fig.~\ref{fig:posteriors}. 
The large covariance between the $^3$H and $^4$He binding energies~\cite{Tjon1975} is seen in the first panel, with the posterior following the trend of the $c_D$-$c_E$ curve found in Ref.~\cite{Gazit2009}, albeit slightly broadened due to the theoretical uncertainty contributions in the denominator. The cross section has a large uncertainty band, particularly around 1.5 MeV, as well as for energies greater than 3.5 MeV.
Inclusion of the charge radius shifts the peak of the posterior towards more negative values, but does little to otherwise restrict the acceptable range of LECs. The cross section is also better constrained at higher energies but the peak still shows a similar variance as before.
Finally, by including the \nalpha scattering resonance positions and widths, we obtain a tighter posterior distribution centered near the optimal values found in the previous work of Ref.~\cite{Gazit2019}. 
The maximum likelihood a posteriori LECs ($c_D =0.925 \pm 0.349$, $c_E=-0.00806 +/- 0.0708$) differ slightly from the values obtained in Ref.~\cite{Gazit2019} and they come with a significantly reduced confidence interval. The marginalization over the entire posterior yields a posterior total \nalpha cross section in remarkably close agreement with experiment (bottom right panel of Fig.~\ref{fig:posteriors}), with virtually all experimental points within the 95\% confidence interval.

\section{Conclusions}
\label{sec:conclusions}
We presented an extensive investigation of uncertainty contributions to \textit{ab initio} calculations of $n$-$\alpha$ scattering owing both to the truncation of  the chiral expansion for the underlying NN+3N Hamiltonian and the uncertainty in the determination of the $c_D$ and $c_E$ LECs of the 3N force. 
For the purpose of such study, we constrained four models of NN+3N chiral Hamiltonians by combining the NN potentials ranging from the third (N$^2$LO) to the fifth order (N$^4$LO) of the chiral expansion of Ref.~\cite{Entem2017} with the leading-order 3N force using a mixture of local and non-local regulators.  The low-energy constants for the contact plus one-pion exchange ($c_D$) and contact ($c_E$) terms of the 3N force were constrained to the binding energy and the $\beta$-decay half life of tritium, taking into account the uncertainty due to the truncation of the chiral expansion.  

The computed $^4$He properties and $n$+$\alpha$ scattering phase shifts demonstrated a rapid order-by-order convergence. Nevertheless, we were unable to accurately reproduce the low-energy $n$-$\alpha$ phase shifts within the parametrization found in Ref.~\cite{Entem2017}, even when accounting for the chiral uncertainty. Given that the chiral expansion is sufficiently converged at N$^4$LO, this discrepancy can either be attributed to deficiencies in the two-body sector of the interaction, or to the omission of higher order components of the 3N force, and in particular those carrying additional spin-orbit strength such as the N$^4$LO 3N contact interactions of Girlanda et al.~\cite{Girlanda2011}. At the same time, such corrections would be unexpectedly large for this order. Closer agreement with empirical data is found when using an older parameterization of the NN interaction at order N$^3$LO~\cite{Entem2003}.

To  gain insight into how uncertainties in the 3N low-energy constants propagate throughout the calculation and determine the Bayesian posterior distribution of these parameters,  we employed Gaussian process models to build fast and accurate statistical emulators for  our  {\em ab initio}  calculations for $^3$H and $^4$He g.s.\ properties and $n$-$\alpha$ scattering observables.  

Working with the parameterization of the NN interaction at order N$^3$LO of Ref.~\cite{Entem2003}, the use of static properties of light nuclei such as the binding energies of $^3$H and $^4$He, as well as the charge radius of $^4$He proved to be insufficient to adequately constrain the posterior distribution for the $c_D$ and $c_E$ LECs, pointing to a large covariance between such observables. On the other hand, the dynamic reaction observables provided the necessary constraints, resulting in a localized posterior,  which in turn leads to tighter confidence intervals for the total \nalpha cross section in excellent agreement with experimental data.
 
This work sets the stage for future investigations on the robustness of NCSMC and, in general, \textit{ab initio} predictions using NN+3N Hamiltonians derived in the framework of chiral EFT. The posterior distribution obtained here, and provided as Supplementary Material, along with the statistical methods demonstrated in this work will be used in the future to provide predictive calculations with quantified uncertainties for low-energy light-nucleus reactions relevant for astrophysics and applications such as the $^6$Li$(n,t)^4$He neutron standard cross section~\cite{Carlson2018,Vorabbi2019} (see also Appendix). The availability of chiral interactions at all (possible) orders, consistently obtained with the same power counting scheme as well as the same cutoff procedures, provides invaluable insight in the chiral convergence which we verified to be a significant source of uncertainty in \textit{ab initio} predictions of scattering observables. 
\begin{acknowledgments}
We thank N. Schunck for useful discussions.
Computing support for this work came from the LLNL institutional Computing Grand Challenge
program. 
Prepared in part by LLNL under Contract DE-AC52-07NA27344. 
This material is based upon work supported by the U.S.\ Department of Energy, Office of Science, Office of Nuclear Physics, 
under Work Proposal No.\ SCW0498, LLNL LDRD projects No. 18-ERD-008, 20-LW-046, and by the NSERC Grant
SAPIN-2016-00033.  
TRIUMF receives federal funding via a contribution agreement with the National Research Council of Canada. 
\end{acknowledgments}

\appendix

\section{Code Runtime Reduction using GPUs}
GPMs help minimizing the number of runs required to carry out uncertainty quantification. 
At the same time, even with a small number of design points, obtaining  the NCSM wave functions (here $^{4,5}$He) used as input for reaction calculations can still be a challenge. 
Graphics processing unit (GPU) accelerators have emerged as the next step in supercomputing architectures, aimed at increasing computational capabilities while reducing the energy (and cost) footprint of high performance computing.
\begin{figure}[b]
\includegraphics[width=0.475\textwidth]{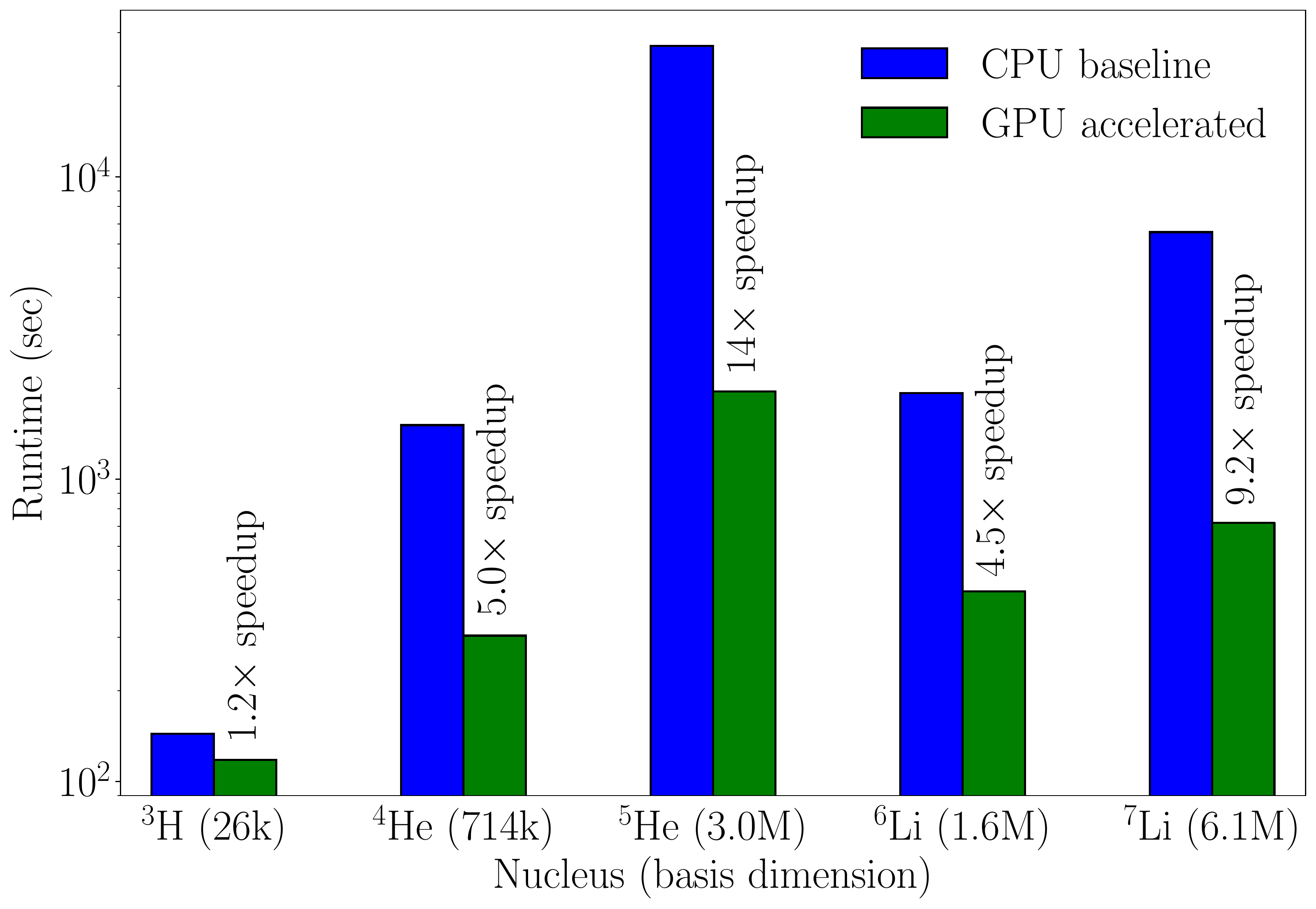}
\caption{Runtime comparison for the CPU and GPU versions of the NCSM diagonalization code (NCSD) within FUSION for various nuclei, performed on the Lassen supercomputer at LLNL. }
\label{fig:GPU}
\end{figure}
The study presented in this work greatly benefitted from efforts to port the FUSION (Fundamental Unified Structure and Interactions of Nuclei) code base for NCSMC calculations to such architectures.
Specifically, we updated the Lanczos algorithm for the NCSM diagonalization by completely offloading the sparse matrix-vector multiplication to a GPU. The computation of a single sparse matrix-vector product showed roughly a 30-50$\times$ speed-up depending on the number of many-body states in the system. To further improve the overall code efficiency by minimizing transfers between central processing unit (CPU) and GPU memory, we allowed the full Hamiltonian matrix to be distributed and stored in GPU memory. 
The total runtimes for various nuclei can be seen in Fig.~\ref{fig:GPU}, along with their respective speedups, obtained when transitioning to the GPU accelerated code. The largest speedup is seen for the $^5$He calculations used in this work, as they were performed for up to 500 Lanczos iterations to obtain the 10 lowest eigenstates.


\begin{thebibliography}{49}%
\makeatletter
\providecommand \@ifxundefined [1]{%
 \@ifx{#1\undefined}
}%
\providecommand \@ifnum [1]{%
 \ifnum #1\expandafter \@firstoftwo
 \else \expandafter \@secondoftwo
 \fi
}%
\providecommand \@ifx [1]{%
 \ifx #1\expandafter \@firstoftwo
 \else \expandafter \@secondoftwo
 \fi
}%
\providecommand \natexlab [1]{#1}%
\providecommand \enquote  [1]{``#1''}%
\providecommand \bibnamefont  [1]{#1}%
\providecommand \bibfnamefont [1]{#1}%
\providecommand \citenamefont [1]{#1}%
\providecommand \href@noop [0]{\@secondoftwo}%
\providecommand \href [0]{\begingroup \@sanitize@url \@href}%
\providecommand \@href[1]{\@@startlink{#1}\@@href}%
\providecommand \@@href[1]{\endgroup#1\@@endlink}%
\providecommand \@sanitize@url [0]{\catcode `\\12\catcode `\$12\catcode
  `\&12\catcode `\#12\catcode `\^12\catcode `\_12\catcode `\%12\relax}%
\providecommand \@@startlink[1]{}%
\providecommand \@@endlink[0]{}%
\providecommand \url  [0]{\begingroup\@sanitize@url \@url }%
\providecommand \@url [1]{\endgroup\@href {#1}{\urlprefix }}%
\providecommand \urlprefix  [0]{URL }%
\providecommand \Eprint [0]{\href }%
\providecommand \doibase [0]{http://dx.doi.org/}%
\providecommand \selectlanguage [0]{\@gobble}%
\providecommand \bibinfo  [0]{\@secondoftwo}%
\providecommand \bibfield  [0]{\@secondoftwo}%
\providecommand \translation [1]{[#1]}%
\providecommand \BibitemOpen [0]{}%
\providecommand \bibitemStop [0]{}%
\providecommand \bibitemNoStop [0]{.\EOS\space}%
\providecommand \EOS [0]{\spacefactor3000\relax}%
\providecommand \BibitemShut  [1]{\csname bibitem#1\endcsname}%
\let\auto@bib@innerbib\@empty
\bibitem [{\citenamefont {Entem}\ \emph {et~al.}(2017)\citenamefont {Entem},
  \citenamefont {Machleidt},\ and\ \citenamefont {Nosyk}}]{Entem2017}%
  \BibitemOpen
  \bibfield  {author} {\bibinfo {author} {\bibfnamefont {D.~R.}\ \bibnamefont
  {Entem}}, \bibinfo {author} {\bibfnamefont {R.}~\bibnamefont {Machleidt}}, \
  and\ \bibinfo {author} {\bibfnamefont {Y.}~\bibnamefont {Nosyk}},\ }\href
  {\doibase 10.1103/PhysRevC.96.024004} {\bibfield  {journal} {\bibinfo
  {journal} {Phys. Rev. C}\ }\textbf {\bibinfo {volume} {96}},\ \bibinfo
  {pages} {024004} (\bibinfo {year} {2017})}\BibitemShut {NoStop}%
\bibitem [{\citenamefont {Gazit}\ \emph {et~al.}(2019)\citenamefont {Gazit},
  \citenamefont {Quaglioni},\ and\ \citenamefont {Navr\'atil}}]{Gazit2019}%
  \BibitemOpen
  \bibfield  {author} {\bibinfo {author} {\bibfnamefont {D.}~\bibnamefont
  {Gazit}}, \bibinfo {author} {\bibfnamefont {S.}~\bibnamefont {Quaglioni}}, \
  and\ \bibinfo {author} {\bibfnamefont {P.}~\bibnamefont {Navr\'atil}},\
  }\href {\doibase 10.1103/PhysRevLett.122.029901} {\bibfield  {journal}
  {\bibinfo  {journal} {Phys. Rev. Lett.}\ }\textbf {\bibinfo {volume} {122}},\
  \bibinfo {pages} {029901} (\bibinfo {year} {2019})}\BibitemShut {NoStop}%
\bibitem [{\citenamefont {Entem}\ and\ \citenamefont
  {Machleidt}(2003)}]{Entem2003}%
  \BibitemOpen
  \bibfield  {author} {\bibinfo {author} {\bibfnamefont {D.~R.}\ \bibnamefont
  {Entem}}\ and\ \bibinfo {author} {\bibfnamefont {R.}~\bibnamefont
  {Machleidt}},\ }\href {\doibase 10.1103/PhysRevC.68.041001} {\bibfield
  {journal} {\bibinfo  {journal} {Phys. Rev. C}\ }\textbf {\bibinfo {volume}
  {68}},\ \bibinfo {pages} {041001} (\bibinfo {year} {2003})}\BibitemShut
  {NoStop}%
\bibitem [{\citenamefont {Baroni}\ \emph
  {et~al.}(2013{\natexlab{a}})\citenamefont {Baroni}, \citenamefont
  {Navr\'{a}til},\ and\ \citenamefont {Quaglioni}}]{Baroni2013}%
  \BibitemOpen
  \bibfield  {author} {\bibinfo {author} {\bibfnamefont {S.}~\bibnamefont
  {Baroni}}, \bibinfo {author} {\bibfnamefont {P.}~\bibnamefont
  {Navr\'{a}til}}, \ and\ \bibinfo {author} {\bibfnamefont {S.}~\bibnamefont
  {Quaglioni}},\ }\href {\doibase 10.1103/PhysRevLett.110.022505} {\bibfield
  {journal} {\bibinfo  {journal} {Phys. Rev. Lett.}\ }\textbf {\bibinfo
  {volume} {110}},\ \bibinfo {pages} {022505} (\bibinfo {year}
  {2013}{\natexlab{a}})}\BibitemShut {NoStop}%
\bibitem [{\citenamefont {Baroni}\ \emph
  {et~al.}(2013{\natexlab{b}})\citenamefont {Baroni}, \citenamefont
  {Navr\'{a}til},\ and\ \citenamefont {Quaglioni}}]{Baroni2013a}%
  \BibitemOpen
  \bibfield  {author} {\bibinfo {author} {\bibfnamefont {S.}~\bibnamefont
  {Baroni}}, \bibinfo {author} {\bibfnamefont {P.}~\bibnamefont
  {Navr\'{a}til}}, \ and\ \bibinfo {author} {\bibfnamefont {S.}~\bibnamefont
  {Quaglioni}},\ }\href {\doibase 10.1103/PhysRevC.87.034326} {\bibfield
  {journal} {\bibinfo  {journal} {Phys. Rev. C}\ }\textbf {\bibinfo {volume}
  {87}},\ \bibinfo {pages} {034326} (\bibinfo {year}
  {2013}{\natexlab{b}})}\BibitemShut {NoStop}%
\bibitem [{\citenamefont {Quaglioni}\ \emph {et~al.}(2018)\citenamefont
  {Quaglioni}, \citenamefont {Romero-Redondo}, \citenamefont {Navr\'atil},\
  and\ \citenamefont {Hupin}}]{Quaglioni2018}%
  \BibitemOpen
  \bibfield  {author} {\bibinfo {author} {\bibfnamefont {S.}~\bibnamefont
  {Quaglioni}}, \bibinfo {author} {\bibfnamefont {C.}~\bibnamefont
  {Romero-Redondo}}, \bibinfo {author} {\bibfnamefont {P.}~\bibnamefont
  {Navr\'atil}}, \ and\ \bibinfo {author} {\bibfnamefont {G.}~\bibnamefont
  {Hupin}},\ }\href {\doibase 10.1103/PhysRevC.97.034332} {\bibfield  {journal}
  {\bibinfo  {journal} {Phys. Rev. C}\ }\textbf {\bibinfo {volume} {97}},\
  \bibinfo {pages} {034332} (\bibinfo {year} {2018})}\BibitemShut {NoStop}%
\bibitem [{\citenamefont {Weinberg}(1990)}]{Weinberg1990}%
  \BibitemOpen
  \bibfield  {author} {\bibinfo {author} {\bibfnamefont {S.}~\bibnamefont
  {Weinberg}},\ }\href {\doibase 10.1016/0370-2693(90)90938-3} {\bibfield
  {journal} {\bibinfo  {journal} {Phys. Lett. B}\ }\textbf {\bibinfo {volume}
  {251}},\ \bibinfo {pages} {288} (\bibinfo {year} {1990})}\BibitemShut
  {NoStop}%
\bibitem [{\citenamefont {Epelbaum}(2006)}]{Epelbaum2006}%
  \BibitemOpen
  \bibfield  {author} {\bibinfo {author} {\bibfnamefont {E.}~\bibnamefont
  {Epelbaum}},\ }\href {\doibase 10.1016/j.ppnp.2005.09.002} {\bibfield
  {journal} {\bibinfo  {journal} {Prog. Part. Nucl. Phys.}\ }\textbf {\bibinfo
  {volume} {57}},\ \bibinfo {pages} {654} (\bibinfo {year} {2006})}\BibitemShut
  {NoStop}%
\bibitem [{\citenamefont {Epelbaum}\ \emph {et~al.}(2009)\citenamefont
  {Epelbaum}, \citenamefont {{H.-W. Hammer}},\ and\ \citenamefont
  {Mei\ss~ner}}]{Epelbaum2009}%
  \BibitemOpen
  \bibfield  {author} {\bibinfo {author} {\bibfnamefont {E.}~\bibnamefont
  {Epelbaum}}, \bibinfo {author} {\bibnamefont {{H.-W. Hammer}}}, \ and\
  \bibinfo {author} {\bibfnamefont {U.-G.}\ \bibnamefont {Mei\ss~ner}},\ }\href
  {\doibase 10.1103/RevModPhys.81.1773} {\bibfield  {journal} {\bibinfo
  {journal} {Rev. Mod. Phys.}\ }\textbf {\bibinfo {volume} {81}},\ \bibinfo
  {pages} {1773} (\bibinfo {year} {2009})}\BibitemShut {NoStop}%
\bibitem [{\citenamefont {Hupin}\ \emph {et~al.}(2019)\citenamefont {Hupin},
  \citenamefont {Quaglioni},\ and\ \citenamefont {Navr{\'a}til}}]{Hupin2019}%
  \BibitemOpen
  \bibfield  {author} {\bibinfo {author} {\bibfnamefont {G.}~\bibnamefont
  {Hupin}}, \bibinfo {author} {\bibfnamefont {S.}~\bibnamefont {Quaglioni}}, \
  and\ \bibinfo {author} {\bibfnamefont {P.}~\bibnamefont {Navr{\'a}til}},\
  }\href {\doibase 10.1038/s41467-018-08052-6} {\bibfield  {journal} {\bibinfo
  {journal} {Nature Communications}\ }\textbf {\bibinfo {volume} {10}},\
  \bibinfo {pages} {351} (\bibinfo {year} {2019})}\BibitemShut {NoStop}%
\bibitem [{\citenamefont {Hupin}\ \emph {et~al.}(2015)\citenamefont {Hupin},
  \citenamefont {Quaglioni},\ and\ \citenamefont {Navr\'atil}}]{Hupin2015}%
  \BibitemOpen
  \bibfield  {author} {\bibinfo {author} {\bibfnamefont {G.}~\bibnamefont
  {Hupin}}, \bibinfo {author} {\bibfnamefont {S.}~\bibnamefont {Quaglioni}}, \
  and\ \bibinfo {author} {\bibfnamefont {P.}~\bibnamefont {Navr\'atil}},\
  }\href {\doibase 10.1103/PhysRevLett.114.212502} {\bibfield  {journal}
  {\bibinfo  {journal} {Phys. Rev. Lett.}\ }\textbf {\bibinfo {volume} {114}},\
  \bibinfo {pages} {212502} (\bibinfo {year} {2015})}\BibitemShut {NoStop}%
\bibitem [{\citenamefont {Ekstr\"om}\ and\ \citenamefont
  {Hagen}(2019)}]{Ekstrom2019}%
  \BibitemOpen
  \bibfield  {author} {\bibinfo {author} {\bibfnamefont {A.}~\bibnamefont
  {Ekstr\"om}}\ and\ \bibinfo {author} {\bibfnamefont {G.}~\bibnamefont
  {Hagen}},\ }\href {\doibase 10.1103/PhysRevLett.123.252501} {\bibfield
  {journal} {\bibinfo  {journal} {Phys. Rev. Lett.}\ }\textbf {\bibinfo
  {volume} {123}},\ \bibinfo {pages} {252501} (\bibinfo {year}
  {2019})}\BibitemShut {NoStop}%
\bibitem [{\citenamefont {McDonnell}\ \emph {et~al.}(2015)\citenamefont
  {McDonnell}, \citenamefont {Schunck}, \citenamefont {Higdon}, \citenamefont
  {Sarich}, \citenamefont {Wild},\ and\ \citenamefont
  {Nazarewicz}}]{McDonnell2015}%
  \BibitemOpen
  \bibfield  {author} {\bibinfo {author} {\bibfnamefont {J.~D.}\ \bibnamefont
  {McDonnell}}, \bibinfo {author} {\bibfnamefont {N.}~\bibnamefont {Schunck}},
  \bibinfo {author} {\bibfnamefont {D.}~\bibnamefont {Higdon}}, \bibinfo
  {author} {\bibfnamefont {J.}~\bibnamefont {Sarich}}, \bibinfo {author}
  {\bibfnamefont {S.~M.}\ \bibnamefont {Wild}}, \ and\ \bibinfo {author}
  {\bibfnamefont {W.}~\bibnamefont {Nazarewicz}},\ }\href {\doibase
  10.1103/PhysRevLett.114.122501} {\bibfield  {journal} {\bibinfo  {journal}
  {Phys. Rev. Lett.}\ }\textbf {\bibinfo {volume} {114}},\ \bibinfo {pages}
  {122501} (\bibinfo {year} {2015})}\BibitemShut {NoStop}%
\bibitem [{\citenamefont {Nollett}\ \emph {et~al.}(2007)\citenamefont
  {Nollett}, \citenamefont {Pieper}, \citenamefont {Wiringa}, \citenamefont
  {Carlson},\ and\ \citenamefont {Hale}}]{Nollett2007}%
  \BibitemOpen
  \bibfield  {author} {\bibinfo {author} {\bibfnamefont {K.~M.}\ \bibnamefont
  {Nollett}}, \bibinfo {author} {\bibfnamefont {S.~C.}\ \bibnamefont {Pieper}},
  \bibinfo {author} {\bibfnamefont {R.~B.}\ \bibnamefont {Wiringa}}, \bibinfo
  {author} {\bibfnamefont {J.}~\bibnamefont {Carlson}}, \ and\ \bibinfo
  {author} {\bibfnamefont {G.~M.}\ \bibnamefont {Hale}},\ }\href {\doibase
  10.1103/PhysRevLett.99.022502} {\bibfield  {journal} {\bibinfo  {journal}
  {Phys. Rev. Lett.}\ }\textbf {\bibinfo {volume} {99}},\ \bibinfo {pages}
  {022502} (\bibinfo {year} {2007})}\BibitemShut {NoStop}%
\bibitem [{\citenamefont {Hupin}\ \emph {et~al.}(2013)\citenamefont {Hupin},
  \citenamefont {Langhammer}, \citenamefont {Navr\'{a}til}, \citenamefont
  {Quaglioni}, \citenamefont {Calci},\ and\ \citenamefont {Roth}}]{Hupin2013}%
  \BibitemOpen
  \bibfield  {author} {\bibinfo {author} {\bibfnamefont {G.}~\bibnamefont
  {Hupin}}, \bibinfo {author} {\bibfnamefont {J.}~\bibnamefont {Langhammer}},
  \bibinfo {author} {\bibfnamefont {P.}~\bibnamefont {Navr\'{a}til}}, \bibinfo
  {author} {\bibfnamefont {S.}~\bibnamefont {Quaglioni}}, \bibinfo {author}
  {\bibfnamefont {A.}~\bibnamefont {Calci}}, \ and\ \bibinfo {author}
  {\bibfnamefont {R.}~\bibnamefont {Roth}},\ }\href {\doibase
  10.1103/PhysRevC.88.054622} {\bibfield  {journal} {\bibinfo  {journal} {Phys.
  Rev. C}\ }\textbf {\bibinfo {volume} {88}},\ \bibinfo {pages} {054622}
  (\bibinfo {year} {2013})}\BibitemShut {NoStop}%
\bibitem [{\citenamefont {Lynn}\ \emph {et~al.}(2016)\citenamefont {Lynn},
  \citenamefont {Tews}, \citenamefont {Carlson}, \citenamefont {Gandolfi},
  \citenamefont {Gezerlis}, \citenamefont {Schmidt},\ and\ \citenamefont
  {Schwenk}}]{Lynn2016}%
  \BibitemOpen
  \bibfield  {author} {\bibinfo {author} {\bibfnamefont {J.~E.}\ \bibnamefont
  {Lynn}}, \bibinfo {author} {\bibfnamefont {I.}~\bibnamefont {Tews}}, \bibinfo
  {author} {\bibfnamefont {J.}~\bibnamefont {Carlson}}, \bibinfo {author}
  {\bibfnamefont {S.}~\bibnamefont {Gandolfi}}, \bibinfo {author}
  {\bibfnamefont {A.}~\bibnamefont {Gezerlis}}, \bibinfo {author}
  {\bibfnamefont {K.~E.}\ \bibnamefont {Schmidt}}, \ and\ \bibinfo {author}
  {\bibfnamefont {A.}~\bibnamefont {Schwenk}},\ }\href {\doibase
  10.1103/PhysRevLett.116.062501} {\bibfield  {journal} {\bibinfo  {journal}
  {Phys. Rev. Lett.}\ }\textbf {\bibinfo {volume} {116}},\ \bibinfo {pages}
  {062501} (\bibinfo {year} {2016})}\BibitemShut {NoStop}%
\bibitem [{\citenamefont {van Kolck}(1994)}]{VanKolck94}%
  \BibitemOpen
  \bibfield  {author} {\bibinfo {author} {\bibfnamefont {U.}~\bibnamefont {van
  Kolck}},\ }\href {\doibase 10.1103/PhysRevC.49.2932} {\bibfield  {journal}
  {\bibinfo  {journal} {Phys. Rev. C}\ }\textbf {\bibinfo {volume} {49}},\
  \bibinfo {pages} {2932} (\bibinfo {year} {1994})}\BibitemShut {NoStop}%
\bibitem [{\citenamefont {Navr\'{a}til}(2007)}]{Navratil2007}%
  \BibitemOpen
  \bibfield  {author} {\bibinfo {author} {\bibfnamefont {P.}~\bibnamefont
  {Navr\'{a}til}},\ }\href {\doibase 10.1007/s00601-007-0193-3} {\bibfield
  {journal} {\bibinfo  {journal} {Few-Body Syst.}\ }\textbf {\bibinfo {volume}
  {41}},\ \bibinfo {pages} {117} (\bibinfo {year} {2007})}\BibitemShut
  {NoStop}%
\bibitem [{\citenamefont {Ekstr\"om}\ \emph {et~al.}(2015)\citenamefont
  {Ekstr\"om}, \citenamefont {Jansen}, \citenamefont {Wendt}, \citenamefont
  {Hagen}, \citenamefont {Papenbrock}, \citenamefont {Carlsson}, \citenamefont
  {Forss\'en}, \citenamefont {Hjorth-Jensen}, \citenamefont {Navr\'atil},\ and\
  \citenamefont {Nazarewicz}}]{EkJa15}%
  \BibitemOpen
  \bibfield  {author} {\bibinfo {author} {\bibfnamefont {A.}~\bibnamefont
  {Ekstr\"om}}, \bibinfo {author} {\bibfnamefont {G.~R.}\ \bibnamefont
  {Jansen}}, \bibinfo {author} {\bibfnamefont {K.~A.}\ \bibnamefont {Wendt}},
  \bibinfo {author} {\bibfnamefont {G.}~\bibnamefont {Hagen}}, \bibinfo
  {author} {\bibfnamefont {T.}~\bibnamefont {Papenbrock}}, \bibinfo {author}
  {\bibfnamefont {B.~D.}\ \bibnamefont {Carlsson}}, \bibinfo {author}
  {\bibfnamefont {C.}~\bibnamefont {Forss\'en}}, \bibinfo {author}
  {\bibfnamefont {M.}~\bibnamefont {Hjorth-Jensen}}, \bibinfo {author}
  {\bibfnamefont {P.}~\bibnamefont {Navr\'atil}}, \ and\ \bibinfo {author}
  {\bibfnamefont {W.}~\bibnamefont {Nazarewicz}},\ }\href {\doibase
  10.1103/PhysRevC.91.051301} {\bibfield  {journal} {\bibinfo  {journal} {Phys.
  Rev. C}\ }\textbf {\bibinfo {volume} {91}},\ \bibinfo {pages} {051301}
  (\bibinfo {year} {2015})}\BibitemShut {NoStop}%
\bibitem [{\citenamefont {Gazit}\ \emph {et~al.}(2009)\citenamefont {Gazit},
  \citenamefont {Quaglioni},\ and\ \citenamefont {Navr\'{a}til}}]{Gazit2009}%
  \BibitemOpen
  \bibfield  {author} {\bibinfo {author} {\bibfnamefont {D.}~\bibnamefont
  {Gazit}}, \bibinfo {author} {\bibfnamefont {S.}~\bibnamefont {Quaglioni}}, \
  and\ \bibinfo {author} {\bibfnamefont {P.}~\bibnamefont {Navr\'{a}til}},\
  }\href {\doibase 10.1103/PhysRevLett.103.102502} {\bibfield  {journal}
  {\bibinfo  {journal} {Phys. Rev. Lett.}\ }\textbf {\bibinfo {volume} {103}},\
  \bibinfo {pages} {102502} (\bibinfo {year} {2009})}\BibitemShut {NoStop}%
\bibitem [{\citenamefont {Hupin}\ \emph {et~al.}(2014)\citenamefont {Hupin},
  \citenamefont {Quaglioni},\ and\ \citenamefont {Navr\'atil}}]{Hupin2014}%
  \BibitemOpen
  \bibfield  {author} {\bibinfo {author} {\bibfnamefont {G.}~\bibnamefont
  {Hupin}}, \bibinfo {author} {\bibfnamefont {S.}~\bibnamefont {Quaglioni}}, \
  and\ \bibinfo {author} {\bibfnamefont {P.}~\bibnamefont {Navr\'atil}},\
  }\href {\doibase 10.1103/PhysRevC.90.061601} {\bibfield  {journal} {\bibinfo
  {journal} {Phys. Rev. C}\ }\textbf {\bibinfo {volume} {90}},\ \bibinfo
  {pages} {061601} (\bibinfo {year} {2014})}\BibitemShut {NoStop}%
\bibitem [{\citenamefont {Calci}\ \emph {et~al.}(2016)\citenamefont {Calci},
  \citenamefont {Navr{\'a}til}, \citenamefont {Roth}, \citenamefont
  {Dohet-Eraly}, \citenamefont {Quaglioni},\ and\ \citenamefont
  {Hupin}}]{Calci2016}%
  \BibitemOpen
  \bibfield  {author} {\bibinfo {author} {\bibfnamefont {A.}~\bibnamefont
  {Calci}}, \bibinfo {author} {\bibfnamefont {P.}~\bibnamefont {Navr{\'a}til}},
  \bibinfo {author} {\bibfnamefont {R.}~\bibnamefont {Roth}}, \bibinfo {author}
  {\bibfnamefont {J.}~\bibnamefont {Dohet-Eraly}}, \bibinfo {author}
  {\bibfnamefont {S.}~\bibnamefont {Quaglioni}}, \ and\ \bibinfo {author}
  {\bibfnamefont {G.}~\bibnamefont {Hupin}},\ }\href {\doibase
  10.1103/PhysRevLett.117.242501} {\bibfield  {journal} {\bibinfo  {journal}
  {Phys. Rev. Lett.}\ }\textbf {\bibinfo {volume} {117}},\ \bibinfo {pages}
  {242501} (\bibinfo {year} {2016})}\BibitemShut {NoStop}%
\bibitem [{\citenamefont {Epelbaum}\ \emph {et~al.}(2015)\citenamefont
  {Epelbaum}, \citenamefont {Krebs},\ and\ \citenamefont
  {Mei{\ss}ner}}]{Epelbaum2015}%
  \BibitemOpen
  \bibfield  {author} {\bibinfo {author} {\bibfnamefont {E.}~\bibnamefont
  {Epelbaum}}, \bibinfo {author} {\bibfnamefont {H.}~\bibnamefont {Krebs}}, \
  and\ \bibinfo {author} {\bibfnamefont {U.~G.}\ \bibnamefont {Mei{\ss}ner}},\
  }\href {\doibase 10.1140/epja/i2015-15053-8} {\bibfield  {journal} {\bibinfo
  {journal} {The European Physical Journal A}\ }\textbf {\bibinfo {volume}
  {51}},\ \bibinfo {pages} {53} (\bibinfo {year} {2015})}\BibitemShut {NoStop}%
\bibitem [{\citenamefont {Furnstahl}\ \emph {et~al.}(2015)\citenamefont
  {Furnstahl}, \citenamefont {Phillips},\ and\ \citenamefont
  {Wesolowski}}]{Furnstahl2015}%
  \BibitemOpen
  \bibfield  {author} {\bibinfo {author} {\bibfnamefont {R.~J.}\ \bibnamefont
  {Furnstahl}}, \bibinfo {author} {\bibfnamefont {D.~R.}\ \bibnamefont
  {Phillips}}, \ and\ \bibinfo {author} {\bibfnamefont {S.}~\bibnamefont
  {Wesolowski}},\ }\href {\doibase 10.1088/0954-3899/42/3/034028} {\bibfield
  {journal} {\bibinfo  {journal} {Journal of Physics G: Nuclear and Particle
  Physics}\ }\textbf {\bibinfo {volume} {42}},\ \bibinfo {pages} {034028}
  (\bibinfo {year} {2015})}\BibitemShut {NoStop}%
\bibitem [{\citenamefont {Melendez}\ \emph {et~al.}(2019)\citenamefont
  {Melendez}, \citenamefont {Furnstahl}, \citenamefont {Phillips},
  \citenamefont {Pratola},\ and\ \citenamefont {Wesolowski}}]{Melendez2019}%
  \BibitemOpen
  \bibfield  {author} {\bibinfo {author} {\bibfnamefont {J.~A.}\ \bibnamefont
  {Melendez}}, \bibinfo {author} {\bibfnamefont {R.~J.}\ \bibnamefont
  {Furnstahl}}, \bibinfo {author} {\bibfnamefont {D.~R.}\ \bibnamefont
  {Phillips}}, \bibinfo {author} {\bibfnamefont {M.~T.}\ \bibnamefont
  {Pratola}}, \ and\ \bibinfo {author} {\bibfnamefont {S.}~\bibnamefont
  {Wesolowski}},\ }\href {\doibase 10.1103/PhysRevC.100.044001} {\bibfield
  {journal} {\bibinfo  {journal} {Phys. Rev. C}\ }\textbf {\bibinfo {volume}
  {100}},\ \bibinfo {pages} {044001} (\bibinfo {year} {2019})}\BibitemShut
  {NoStop}%
\bibitem [{\citenamefont {Baroni}\ \emph {et~al.}(2016)\citenamefont {Baroni},
  \citenamefont {Girlanda}, \citenamefont {Kievsky}, \citenamefont {Marcucci},
  \citenamefont {Schiavilla},\ and\ \citenamefont {Viviani}}]{Baroni2016}%
  \BibitemOpen
  \bibfield  {author} {\bibinfo {author} {\bibfnamefont {A.}~\bibnamefont
  {Baroni}}, \bibinfo {author} {\bibfnamefont {L.}~\bibnamefont {Girlanda}},
  \bibinfo {author} {\bibfnamefont {A.}~\bibnamefont {Kievsky}}, \bibinfo
  {author} {\bibfnamefont {L.~E.}\ \bibnamefont {Marcucci}}, \bibinfo {author}
  {\bibfnamefont {R.}~\bibnamefont {Schiavilla}}, \ and\ \bibinfo {author}
  {\bibfnamefont {M.}~\bibnamefont {Viviani}},\ }\href {\doibase
  10.1103/PhysRevC.94.024003} {\bibfield  {journal} {\bibinfo  {journal} {Phys.
  Rev. C}\ }\textbf {\bibinfo {volume} {94}},\ \bibinfo {pages} {024003}
  (\bibinfo {year} {2016})}\BibitemShut {NoStop}%
\bibitem [{\citenamefont {Navr\'{a}til}\ \emph {et~al.}(2000)\citenamefont
  {Navr\'{a}til}, \citenamefont {Vary},\ and\ \citenamefont
  {Barrett}}]{Navratil2000a}%
  \BibitemOpen
  \bibfield  {author} {\bibinfo {author} {\bibfnamefont {P.}~\bibnamefont
  {Navr\'{a}til}}, \bibinfo {author} {\bibfnamefont {J.~P.}\ \bibnamefont
  {Vary}}, \ and\ \bibinfo {author} {\bibfnamefont {B.~R.}\ \bibnamefont
  {Barrett}},\ }\href {\doibase 10.1103/PhysRevLett.84.5728} {\bibfield
  {journal} {\bibinfo  {journal} {Phys. Rev. Lett.}\ }\textbf {\bibinfo
  {volume} {84}},\ \bibinfo {pages} {5728} (\bibinfo {year}
  {2000})}\BibitemShut {NoStop}%
\bibitem [{\citenamefont {Barrett}\ \emph {et~al.}(2013)\citenamefont
  {Barrett}, \citenamefont {Navr\'{a}til},\ and\ \citenamefont
  {Vary}}]{Barrett2013}%
  \BibitemOpen
  \bibfield  {author} {\bibinfo {author} {\bibfnamefont {B.~R.}\ \bibnamefont
  {Barrett}}, \bibinfo {author} {\bibfnamefont {P.}~\bibnamefont
  {Navr\'{a}til}}, \ and\ \bibinfo {author} {\bibfnamefont {J.~P.}\
  \bibnamefont {Vary}},\ }\href {\doibase 10.1016/j.ppnp.2012.10.003}
  {\bibfield  {journal} {\bibinfo  {journal} {Prog. Part. Nucl. Phys.}\
  }\textbf {\bibinfo {volume} {69}},\ \bibinfo {pages} {131} (\bibinfo {year}
  {2013})}\BibitemShut {NoStop}%
\bibitem [{\citenamefont {Hesse}\ \emph {et~al.}(1998)\citenamefont {Hesse},
  \citenamefont {Sparenberg}, \citenamefont {{Van Raemdonck}},\ and\
  \citenamefont {Baye}}]{Hesse1998}%
  \BibitemOpen
  \bibfield  {author} {\bibinfo {author} {\bibfnamefont {M.}~\bibnamefont
  {Hesse}}, \bibinfo {author} {\bibfnamefont {J.-M.}\ \bibnamefont
  {Sparenberg}}, \bibinfo {author} {\bibfnamefont {F.}~\bibnamefont {{Van
  Raemdonck}}}, \ and\ \bibinfo {author} {\bibfnamefont {D.}~\bibnamefont
  {Baye}},\ }\href {\doibase 10.1016/S0375-9474(98)00435-7} {\bibfield
  {journal} {\bibinfo  {journal} {Nucl. Phys. A}\ }\textbf {\bibinfo {volume}
  {640}},\ \bibinfo {pages} {37} (\bibinfo {year} {1998})}\BibitemShut
  {NoStop}%
\bibitem [{\citenamefont {Romero-Redondo}\ \emph {et~al.}(2016)\citenamefont
  {Romero-Redondo}, \citenamefont {Quaglioni}, \citenamefont {Navr\'atil},\
  and\ \citenamefont {Hupin}}]{Romero-Redondo2016}%
  \BibitemOpen
  \bibfield  {author} {\bibinfo {author} {\bibfnamefont {C.}~\bibnamefont
  {Romero-Redondo}}, \bibinfo {author} {\bibfnamefont {S.}~\bibnamefont
  {Quaglioni}}, \bibinfo {author} {\bibfnamefont {P.}~\bibnamefont
  {Navr\'atil}}, \ and\ \bibinfo {author} {\bibfnamefont {G.}~\bibnamefont
  {Hupin}},\ }\href {\doibase 10.1103/PhysRevLett.117.222501} {\bibfield
  {journal} {\bibinfo  {journal} {Phys. Rev. Lett.}\ }\textbf {\bibinfo
  {volume} {117}},\ \bibinfo {pages} {222501} (\bibinfo {year}
  {2016})}\BibitemShut {NoStop}%
\bibitem [{\citenamefont {Navr{\'{a}}til}\ \emph {et~al.}(2016)\citenamefont
  {Navr{\'{a}}til}, \citenamefont {Quaglioni}, \citenamefont {Hupin},
  \citenamefont {Romero-Redondo},\ and\ \citenamefont {Calci}}]{Navratil2016}%
  \BibitemOpen
  \bibfield  {author} {\bibinfo {author} {\bibfnamefont {P.}~\bibnamefont
  {Navr{\'{a}}til}}, \bibinfo {author} {\bibfnamefont {S.}~\bibnamefont
  {Quaglioni}}, \bibinfo {author} {\bibfnamefont {G.}~\bibnamefont {Hupin}},
  \bibinfo {author} {\bibfnamefont {C.}~\bibnamefont {Romero-Redondo}}, \ and\
  \bibinfo {author} {\bibfnamefont {A.}~\bibnamefont {Calci}},\ }\href
  {\doibase 10.1088/0031-8949/91/5/053002} {\bibfield  {journal} {\bibinfo
  {journal} {Physica Scripta}\ }\textbf {\bibinfo {volume} {91}},\ \bibinfo
  {pages} {053002} (\bibinfo {year} {2016})}\BibitemShut {NoStop}%
\bibitem [{\citenamefont {Gu}\ \emph {et~al.}(2018)\citenamefont {Gu},
  \citenamefont {Wang}, \citenamefont {Berger} \emph {et~al.}}]{gu2018robust}%
  \BibitemOpen
  \bibfield  {author} {\bibinfo {author} {\bibfnamefont {M.}~\bibnamefont
  {Gu}}, \bibinfo {author} {\bibfnamefont {X.}~\bibnamefont {Wang}}, \bibinfo
  {author} {\bibfnamefont {J.~O.}\ \bibnamefont {Berger}},  \emph {et~al.},\
  }\href@noop {} {\bibfield  {journal} {\bibinfo  {journal} {The Annals of
  Statistics}\ }\textbf {\bibinfo {volume} {46}},\ \bibinfo {pages} {3038}
  (\bibinfo {year} {2018})}\BibitemShut {NoStop}%
\bibitem [{\citenamefont {Gu}\ \emph {et~al.}(2019)\citenamefont {Gu},
  \citenamefont {Palomo},\ and\ \citenamefont {Berger}}]{GaSPpackage}%
  \BibitemOpen
  \bibfield  {author} {\bibinfo {author} {\bibfnamefont {M.}~\bibnamefont
  {Gu}}, \bibinfo {author} {\bibfnamefont {J.}~\bibnamefont {Palomo}}, \ and\
  \bibinfo {author} {\bibfnamefont {J.}~\bibnamefont {Berger}},\ }\href
  {https://CRAN.R-project.org/package=RobustGaSP} {\emph {\bibinfo {title}
  {RobustGaSP: Robust Gaussian Stochastic Process Emulation}}} (\bibinfo {year}
  {2019}),\ \bibinfo {note} {r package version 0.5.7}\BibitemShut {NoStop}%
\bibitem [{\citenamefont {Williams}\ and\ \citenamefont
  {Rasmussen}(2006)}]{williams2006}%
  \BibitemOpen
  \bibfield  {author} {\bibinfo {author} {\bibfnamefont {C.~K.}\ \bibnamefont
  {Williams}}\ and\ \bibinfo {author} {\bibfnamefont {C.~E.}\ \bibnamefont
  {Rasmussen}},\ }\href@noop {} {\emph {\bibinfo {title} {Gaussian processes
  for machine learning}}},\ Vol.~\bibinfo {volume} {2}\ (\bibinfo  {publisher}
  {MIT press Cambridge, MA},\ \bibinfo {year} {2006})\BibitemShut {NoStop}%
\bibitem [{\citenamefont {Jurgenson}\ \emph {et~al.}(2011)\citenamefont
  {Jurgenson}, \citenamefont {Navr\'{a}til},\ and\ \citenamefont
  {Furnstahl}}]{Jurgenson2011}%
  \BibitemOpen
  \bibfield  {author} {\bibinfo {author} {\bibfnamefont {E.~D.}\ \bibnamefont
  {Jurgenson}}, \bibinfo {author} {\bibfnamefont {P.}~\bibnamefont
  {Navr\'{a}til}}, \ and\ \bibinfo {author} {\bibfnamefont {R.~J.}\
  \bibnamefont {Furnstahl}},\ }\href {\doibase 10.1103/PhysRevC.83.034301}
  {\bibfield  {journal} {\bibinfo  {journal} {Phys. Rev. C}\ }\textbf {\bibinfo
  {volume} {83}},\ \bibinfo {pages} {034301} (\bibinfo {year}
  {2011})}\BibitemShut {NoStop}%
\bibitem [{\citenamefont {Jurgenson}\ \emph {et~al.}(2013)\citenamefont
  {Jurgenson}, \citenamefont {Maris}, \citenamefont {Furnstahl}, \citenamefont
  {Navr\'{a}til}, \citenamefont {Ormand},\ and\ \citenamefont
  {Vary}}]{Jurgenson2013}%
  \BibitemOpen
  \bibfield  {author} {\bibinfo {author} {\bibfnamefont {E.~D.}\ \bibnamefont
  {Jurgenson}}, \bibinfo {author} {\bibfnamefont {P.}~\bibnamefont {Maris}},
  \bibinfo {author} {\bibfnamefont {R.~J.}\ \bibnamefont {Furnstahl}}, \bibinfo
  {author} {\bibfnamefont {P.}~\bibnamefont {Navr\'{a}til}}, \bibinfo {author}
  {\bibfnamefont {W.~E.}\ \bibnamefont {Ormand}}, \ and\ \bibinfo {author}
  {\bibfnamefont {J.~P.}\ \bibnamefont {Vary}},\ }\href {\doibase
  10.1103/PhysRevC.87.054312} {\bibfield  {journal} {\bibinfo  {journal} {Phys.
  Rev. C}\ }\textbf {\bibinfo {volume} {87}},\ \bibinfo {pages} {054312}
  (\bibinfo {year} {2013})}\BibitemShut {NoStop}%
\bibitem [{\citenamefont {Angeli}\ and\ \citenamefont
  {Marinova}(2013)}]{Angeli2013}%
  \BibitemOpen
  \bibfield  {author} {\bibinfo {author} {\bibfnamefont {I.}~\bibnamefont
  {Angeli}}\ and\ \bibinfo {author} {\bibfnamefont {K.}~\bibnamefont
  {Marinova}},\ }\href {\doibase https://doi.org/10.1016/j.adt.2011.12.006}
  {\bibfield  {journal} {\bibinfo  {journal} {Atomic Data and Nuclear Data
  Tables}\ }\textbf {\bibinfo {volume} {99}},\ \bibinfo {pages} {69 } (\bibinfo
  {year} {2013})}\BibitemShut {NoStop}%
\bibitem [{\citenamefont {Tiesinga}\ \emph {et~al.}(2020)\citenamefont
  {Tiesinga}, \citenamefont {Mohr}, \citenamefont {Newell},\ and\ \citenamefont
  {Taylor}}]{Tiesinga2020}%
  \BibitemOpen
  \bibfield  {author} {\bibinfo {author} {\bibfnamefont {E.}~\bibnamefont
  {Tiesinga}}, \bibinfo {author} {\bibfnamefont {P.~J.}\ \bibnamefont {Mohr}},
  \bibinfo {author} {\bibfnamefont {D.~B.}\ \bibnamefont {Newell}}, \ and\
  \bibinfo {author} {\bibfnamefont {B.~N.}\ \bibnamefont {Taylor}},\ }\href
  {http://physics.nist.gov/constants} {\bibfield  {journal} {\bibinfo
  {journal} {The 2018 CODATA Recommended Values of the Fundamental Physical
  Constants}\ } (\bibinfo {year} {2020})}\BibitemShut {NoStop}%
\bibitem [{\citenamefont {Hale}(2008)}]{Hale}%
  \BibitemOpen
  \bibfield  {author} {\bibinfo {author} {\bibfnamefont {G.~M.}\ \bibnamefont
  {Hale}},\ }\href@noop {} {\bibfield  {journal} {\bibinfo  {journal} {Private
  Communication}\ } (\bibinfo {year} {2008})}\BibitemShut {NoStop}%
\bibitem [{\citenamefont {Girlanda}\ \emph {et~al.}(2011)\citenamefont
  {Girlanda}, \citenamefont {Kievsky},\ and\ \citenamefont
  {Viviani}}]{Girlanda2011}%
  \BibitemOpen
  \bibfield  {author} {\bibinfo {author} {\bibfnamefont {L.}~\bibnamefont
  {Girlanda}}, \bibinfo {author} {\bibfnamefont {A.}~\bibnamefont {Kievsky}}, \
  and\ \bibinfo {author} {\bibfnamefont {M.}~\bibnamefont {Viviani}},\ }\href
  {\doibase 10.1103/PhysRevC.84.014001} {\bibfield  {journal} {\bibinfo
  {journal} {Phys. Rev. C}\ }\textbf {\bibinfo {volume} {84}},\ \bibinfo
  {pages} {014001} (\bibinfo {year} {2011})}\BibitemShut {NoStop}%
\bibitem [{\citenamefont {Santner}\ \emph {et~al.}(2003)\citenamefont
  {Santner}, \citenamefont {Williams}, \citenamefont {Notz},\ and\
  \citenamefont {Williams}}]{santner2003}%
  \BibitemOpen
  \bibfield  {author} {\bibinfo {author} {\bibfnamefont {T.~J.}\ \bibnamefont
  {Santner}}, \bibinfo {author} {\bibfnamefont {B.~J.}\ \bibnamefont
  {Williams}}, \bibinfo {author} {\bibfnamefont {W.}~\bibnamefont {Notz}}, \
  and\ \bibinfo {author} {\bibfnamefont {B.~J.}\ \bibnamefont {Williams}},\
  }\href@noop {} {\emph {\bibinfo {title} {The design and analysis of computer
  experiments}}},\ Vol.~\bibinfo {volume} {1}\ (\bibinfo  {publisher}
  {Springer},\ \bibinfo {year} {2003})\BibitemShut {NoStop}%
\bibitem [{\citenamefont {Kirkpatrick}\ \emph {et~al.}(1983)\citenamefont
  {Kirkpatrick}, \citenamefont {Gelatt},\ and\ \citenamefont
  {Vecchi}}]{kirkpatrick1983}%
  \BibitemOpen
  \bibfield  {author} {\bibinfo {author} {\bibfnamefont {S.}~\bibnamefont
  {Kirkpatrick}}, \bibinfo {author} {\bibfnamefont {C.~D.}\ \bibnamefont
  {Gelatt}}, \ and\ \bibinfo {author} {\bibfnamefont {M.~P.}\ \bibnamefont
  {Vecchi}},\ }\href@noop {} {\bibfield  {journal} {\bibinfo  {journal}
  {science}\ }\textbf {\bibinfo {volume} {220}},\ \bibinfo {pages} {671}
  (\bibinfo {year} {1983})}\BibitemShut {NoStop}%
\bibitem [{\citenamefont {Johnson}\ \emph {et~al.}(1990)\citenamefont
  {Johnson}, \citenamefont {Moore},\ and\ \citenamefont
  {Ylvisaker}}]{johnson1990}%
  \BibitemOpen
  \bibfield  {author} {\bibinfo {author} {\bibfnamefont {M.~E.}\ \bibnamefont
  {Johnson}}, \bibinfo {author} {\bibfnamefont {L.~M.}\ \bibnamefont {Moore}},
  \ and\ \bibinfo {author} {\bibfnamefont {D.}~\bibnamefont {Ylvisaker}},\
  }\href@noop {} {\bibfield  {journal} {\bibinfo  {journal} {Journal of
  statistical planning and inference}\ }\textbf {\bibinfo {volume} {26}},\
  \bibinfo {pages} {131} (\bibinfo {year} {1990})}\BibitemShut {NoStop}%
\bibitem [{\citenamefont {Dohet-Eraly}\ \emph {et~al.}(2016)\citenamefont
  {Dohet-Eraly}, \citenamefont {Navr{\'a}til}, \citenamefont {Quaglioni},
  \citenamefont {Horiuchi}, \citenamefont {Hupin},\ and\ \citenamefont
  {Raimondi}}]{DohetEraly2016}%
  \BibitemOpen
  \bibfield  {author} {\bibinfo {author} {\bibfnamefont {J.}~\bibnamefont
  {Dohet-Eraly}}, \bibinfo {author} {\bibfnamefont {P.}~\bibnamefont
  {Navr{\'a}til}}, \bibinfo {author} {\bibfnamefont {S.}~\bibnamefont
  {Quaglioni}}, \bibinfo {author} {\bibfnamefont {W.}~\bibnamefont {Horiuchi}},
  \bibinfo {author} {\bibfnamefont {G.}~\bibnamefont {Hupin}}, \ and\ \bibinfo
  {author} {\bibfnamefont {F.}~\bibnamefont {Raimondi}},\ }\href {\doibase
  https://doi.org/10.1016/j.physletb.2016.04.021} {\bibfield  {journal}
  {\bibinfo  {journal} {Physics Letters B}\ }\textbf {\bibinfo {volume}
  {757}},\ \bibinfo {pages} {430 } (\bibinfo {year} {2016})}\BibitemShut
  {NoStop}%
\bibitem [{\citenamefont {Brown}\ and\ \citenamefont
  {Richter}(2006)}]{Brown2006}%
  \BibitemOpen
  \bibfield  {author} {\bibinfo {author} {\bibfnamefont {B.~A.}\ \bibnamefont
  {Brown}}\ and\ \bibinfo {author} {\bibfnamefont {W.~A.}\ \bibnamefont
  {Richter}},\ }\href {\doibase 10.1103/PhysRevC.74.034315} {\bibfield
  {journal} {\bibinfo  {journal} {Phys. Rev. C}\ }\textbf {\bibinfo {volume}
  {74}},\ \bibinfo {pages} {034315} (\bibinfo {year} {2006})}\BibitemShut
  {NoStop}%
\bibitem [{\citenamefont {Haario}\ \emph {et~al.}(2006)\citenamefont {Haario},
  \citenamefont {Laine}, \citenamefont {Mira},\ and\ \citenamefont
  {Saksman}}]{haario2006dram}%
  \BibitemOpen
  \bibfield  {author} {\bibinfo {author} {\bibfnamefont {H.}~\bibnamefont
  {Haario}}, \bibinfo {author} {\bibfnamefont {M.}~\bibnamefont {Laine}},
  \bibinfo {author} {\bibfnamefont {A.}~\bibnamefont {Mira}}, \ and\ \bibinfo
  {author} {\bibfnamefont {E.}~\bibnamefont {Saksman}},\ }\href@noop {}
  {\bibfield  {journal} {\bibinfo  {journal} {Statistics and computing}\
  }\textbf {\bibinfo {volume} {16}},\ \bibinfo {pages} {339} (\bibinfo {year}
  {2006})}\BibitemShut {NoStop}%
\bibitem [{\citenamefont {Tjon}(1975)}]{Tjon1975}%
  \BibitemOpen
  \bibfield  {author} {\bibinfo {author} {\bibfnamefont {J.}~\bibnamefont
  {Tjon}},\ }\href {\doibase https://doi.org/10.1016/0370-2693(75)90378-0}
  {\bibfield  {journal} {\bibinfo  {journal} {Physics Letters B}\ }\textbf
  {\bibinfo {volume} {56}},\ \bibinfo {pages} {217 } (\bibinfo {year}
  {1975})}\BibitemShut {NoStop}%
\bibitem [{\citenamefont {Carlson}\ \emph {et~al.}(2018)\citenamefont
  {Carlson}, \citenamefont {Pronyaev}, \citenamefont {Capote}, \citenamefont
  {Hale}, \citenamefont {Chen}, \citenamefont {Duran}, \citenamefont {Hambsch},
  \citenamefont {Kunieda}, \citenamefont {Mannhart}, \citenamefont
  {Marcinkevicius}, \citenamefont {Nelson}, \citenamefont {Neudecker},
  \citenamefont {Noguere}, \citenamefont {Paris}, \citenamefont {Simakov},
  \citenamefont {Schillebeeckx}, \citenamefont {Smith}, \citenamefont {Tao},
  \citenamefont {Trkov}, \citenamefont {Wallner},\ and\ \citenamefont
  {Wang}}]{Carlson2018}%
  \BibitemOpen
  \bibfield  {author} {\bibinfo {author} {\bibfnamefont {A.}~\bibnamefont
  {Carlson}}, \bibinfo {author} {\bibfnamefont {V.}~\bibnamefont {Pronyaev}},
  \bibinfo {author} {\bibfnamefont {R.}~\bibnamefont {Capote}}, \bibinfo
  {author} {\bibfnamefont {G.}~\bibnamefont {Hale}}, \bibinfo {author}
  {\bibfnamefont {Z.-P.}\ \bibnamefont {Chen}}, \bibinfo {author}
  {\bibfnamefont {I.}~\bibnamefont {Duran}}, \bibinfo {author} {\bibfnamefont
  {F.-J.}\ \bibnamefont {Hambsch}}, \bibinfo {author} {\bibfnamefont
  {S.}~\bibnamefont {Kunieda}}, \bibinfo {author} {\bibfnamefont
  {W.}~\bibnamefont {Mannhart}}, \bibinfo {author} {\bibfnamefont
  {B.}~\bibnamefont {Marcinkevicius}}, \bibinfo {author} {\bibfnamefont
  {R.}~\bibnamefont {Nelson}}, \bibinfo {author} {\bibfnamefont
  {D.}~\bibnamefont {Neudecker}}, \bibinfo {author} {\bibfnamefont
  {G.}~\bibnamefont {Noguere}}, \bibinfo {author} {\bibfnamefont
  {M.}~\bibnamefont {Paris}}, \bibinfo {author} {\bibfnamefont
  {S.}~\bibnamefont {Simakov}}, \bibinfo {author} {\bibfnamefont
  {P.}~\bibnamefont {Schillebeeckx}}, \bibinfo {author} {\bibfnamefont
  {D.}~\bibnamefont {Smith}}, \bibinfo {author} {\bibfnamefont
  {X.}~\bibnamefont {Tao}}, \bibinfo {author} {\bibfnamefont {A.}~\bibnamefont
  {Trkov}}, \bibinfo {author} {\bibfnamefont {A.}~\bibnamefont {Wallner}}, \
  and\ \bibinfo {author} {\bibfnamefont {W.}~\bibnamefont {Wang}},\ }\href
  {\doibase https://doi.org/10.1016/j.nds.2018.02.002} {\bibfield  {journal}
  {\bibinfo  {journal} {Nuclear Data Sheets}\ }\textbf {\bibinfo {volume}
  {148}},\ \bibinfo {pages} {143 } (\bibinfo {year} {2018})},\ \bibinfo {note}
  {special Issue on Nuclear Reaction Data}\BibitemShut {NoStop}%
\bibitem [{\citenamefont {Vorabbi}\ \emph {et~al.}(2019)\citenamefont
  {Vorabbi}, \citenamefont {Navr\'atil}, \citenamefont {Quaglioni},\ and\
  \citenamefont {Hupin}}]{Vorabbi2019}%
  \BibitemOpen
  \bibfield  {author} {\bibinfo {author} {\bibfnamefont {M.}~\bibnamefont
  {Vorabbi}}, \bibinfo {author} {\bibfnamefont {P.}~\bibnamefont {Navr\'atil}},
  \bibinfo {author} {\bibfnamefont {S.}~\bibnamefont {Quaglioni}}, \ and\
  \bibinfo {author} {\bibfnamefont {G.}~\bibnamefont {Hupin}},\ }\href
  {\doibase 10.1103/PhysRevC.100.024304} {\bibfield  {journal} {\bibinfo
  {journal} {Phys. Rev. C}\ }\textbf {\bibinfo {volume} {100}},\ \bibinfo
  {pages} {024304} (\bibinfo {year} {2019})}\BibitemShut {NoStop}%
\end{thebibliography}
\end{document}